# Structure et propriétés des verres de silicate de plomb

Daniel Caurant[(1)], Gilles Wallez[(1,2)], Odile Majérus[(1)],

Gauthier Roisine[(1)], Thibault Charpentier[(3)]

[(1)]*Chimie ParisTech, Université PSL, CNRS, Institut de Recherche de Chimie Paris (IRCP), UMR 8247, 11 rue P. et M. Curie, 75005 Paris, France*
[(2)]*Sorbonne Université, UFR 926, 75005 Paris, France.*
[(3)]*Université Paris-Saclay, CEA, CNRS, NIMBE, 91191 Gif-sur-Yvette Cedex, France*

## 3.1. Introduction

Comme nous l'avons vu dans le Chapitre 1, en raison de leurs propriétés particulières, le domaine d'application des verres renfermant de l'oxyde de plomb est très vaste. Il s'étend des verres du monde de la cristallerie, des émaux et glaçures, à celui des verres d'optique à haut indice de réfraction, en passant par les verres absorbant les rayonnements ionisants et les verres destinés à réaliser des soudures à basse température entre métaux, céramiques ou verres (Frieser 1975 ; Rabinovich 1976 ; Takamori 1979 ; Volf 1984). Pour ces différentes applications, la composition du verre doit bien sûr être adaptée tant en ce qui concerne son élaboration et sa mise en forme (viscosité, coefficient de dilatation, température de transition vitreuse $T_g$) que ses propriétés dans les conditions d'utilisation (indice de réfraction, transparence, couleur, durabilité chimique, résistivité électrique…). Ces différentes propriétés qui concernent aussi bien le liquide au cours de sa fusion, de sa mise en forme et de son refroidissement (trempe) que le verre final, c'est-à-dire le solide amorphe obtenu après trempe du liquide surfondu en dessous de $T_g$, sont étroitement liées à sa structure. Dans le cas des verres d'oxydes qui nous intéressent ici, cela correspond à la façon dont les polyèdres associés aux différentes espèces cationiques présentes entourées d'anions oxygène $O^{2-}$ sont agencés les uns par rapport aux autres au sein du réseau qui constitue la structure du verre, ce réseau ne présentant pas d'ordre à grande distance à la différence des cristaux (figure 3.1). Dans le cas des verres renfermant du plomb, il est donc particulièrement important de comprendre la façon dont les ions $Pb^{2+}$ vont s'insérer dans la structure de ce réseau en fonction de la teneur en oxyde de plomb et de la composition du verre (nature de l'oxyde formateur ($SiO_2$, $B_2O_3$, $P_2O_5$), effet de la présence d'oxydes modificateurs (oxydes alcalins et alcalino-terreux) et intermédiaires ($Al_2O_3$))[1] et d'évaluer leur impact sur la structure de ce réseau ainsi que sur l'environnement et la distribution d'espèces particulières pouvant être présentes telles que les ions de métaux de transition utilisés pour la coloration (voir Chapitre 4).

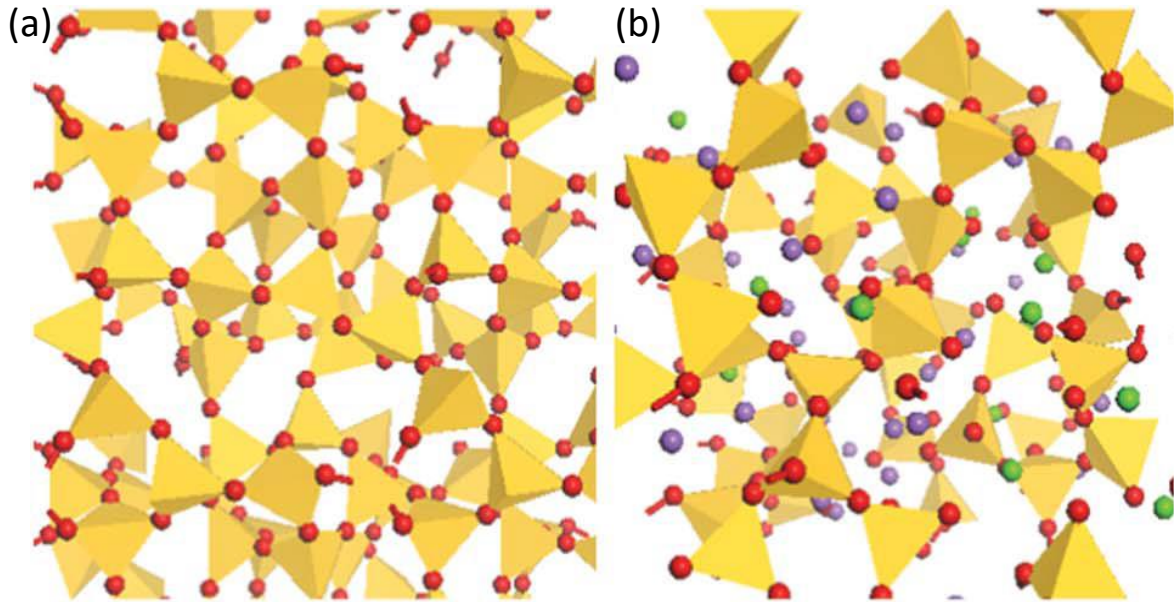

**Figure 3.1.** *Exemple de représentation tridimensionnelle du verre de silice pure (a) montrant les tétraèdres $SiO_4$ (en jaune) tous connectés entre eux par des atomes d'oxygène pontants (en rouge) et (b) d'un verre silico-sodo-calcique ($SiO_2$-$Na_2O$-$CaO$) montrant la coupure du réseau silicaté avec la formation d'atomes d'oxygène non-pontants connectés à un seul tétraèdre $SiO_4$. Les ions modificateurs $Na^+$ (en violet) et*

---

[1] Un oxyde formateur peut se définir comme étant un oxyde qui seul par fusion + refroidissement plus ou moins rapide (trempe) conduit facilement à un verre du fait notamment de la viscosité élevée du liquide correspondant au voisinage de la température de fusion $T_f$ et dans son état surfondu ($T < T_f$). Un verre peut quant à lui être défini comme un solide ne présentant pas d'ordre à grande distance ($\geq 10$Å) au niveau de son organisation atomique (à la différence des matériaux cristallisés) mais présentant une température de transition vitreuse $T_g$ (zone de température où le liquide se fige au refroidissement et où le verre tend à commencer à se ramollir au chauffage). On peut considérer un verre comme un liquide surfondu figé lors de la trempe. Par rapport aux matériaux cristallisés les verres présentent donc un excès d'énergie libre et sont dans un état hors équilibre du point de vue de la thermodynamique (lors du chauffage au-dessus de $T_g$ ils tendent à cristalliser). Les principaux oxydes formateurs sont $SiO_2$, $B_2O_3$, $P_2O_5$ et $GeO_2$ qui conduisent facilement à des verres et dont la structure est constituée d'un réseau continu aléatoire de tétraèdres ($SiO_4$, $PO_4$, $GeO_4$) ou de triangles ($BO_3$) connectés uniquement par les sommets (faible excès d'énergie libre par rapport à l'état ordonné cristallisé), au moyen de liaisons directionnelles à caractère iono-covalent (modèle de Zachariasen (Zachariasen 1932)) (figure3.1a). Un oxyde modificateur peut quant à lui être défini comme un oxyde qui seul ne conduit jamais à un verre par fusion + trempe (même extrêmement rapide) du liquide en fusion, du fait notamment de la viscosité trop faible du liquide et de l'organisation structurale des polyèdres les uns par rapport aux autres difficile à maintenir dans un état désordonné (excès d'énergie libre entre état désordonné et ordonné bien plus important que pour les oxydes formateurs). On rajoute cependant aux oxydes formateurs des quantités plus ou moins importantes d'oxydes modificateurs (oxydes alcalins et alcalino-terreux notamment) dans les verres afin de faciliter la fusion du mélange (rôle de fondant), l'affinage du liquide (élimination des microbulles et homogénéisation) et le formage des objets (rôle de fluidifiant), ainsi que pour agir sur les propriétés du verre final. Au niveau structural, les oxydes modificateurs (présents sous forme de cations) jouent au moyen de liaisons non-directionnelles à caractère ionique un rôle de compensateur au voisinage des charges négatives présentes dans la structure telles que celles produites par la coupure de liaisons iono-covalentes comme les liaisons Si-O-Si (formation d'atomes d'oxygène non-pontants, $Si-O^-$) (figure 3.1b) ou celles dues à la présence d'espèces chargées négativement telles que $AlO_4^-$ ou $BO_4^-$. Comme les oxydes modificateurs, les oxydes intermédiaires - tel que $Al_2O_3$ - ne conduisent jamais seuls à des verres par fusion + trempe, mais au niveau structural ils jouent généralement un rôle différent des modificateurs et peuvent rentrer dans la structure du réseau formateur. Signalons quand même qu'en l'absence d'oxyde formateur et en présence d'un ou plusieurs oxydes modificateurs, les oxydes intermédiaires peuvent également conduire dans certains cas à des verres comme c'est le cas des verres d'aluminates (Pye 1988).

*Ca²⁺ (en vert) compensent la charge négative des atomes d'oxygène non-pontants. Reproduit à partir de (Charpentier et al. 2013) avec la permission de The Royal Society of Chemistry.*

Dans le présent chapitre nous nous focaliserons tout d'abord sur la chimie du plomb et plus particulièrement sur les propriétés et la stéréochimie de l'ion $Pb^{2+}$ (espèce sous laquelle est présent le plomb dans les verres d'oxydes), la nature de la liaison chimique Pb-O et la structure cristalline de l'oxyde PbO pur (stéréochimie et mode de connexion des polyèdres). Puis, comme les verres d'oxydes à base de plomb d'intérêt sont très majoritairement des verres silicatés, nous nous intéresserons à la façon dont les ions $Pb^{2+}$ s'insèrent dans le réseau vitreux de la silice et modifient sa structure d'abord dans le cas du système binaire simple $SiO_2$-PbO (pour ce système l'évolution de la structure des phases cristallines binaires sera également abordée en fonction de leur teneur en PbO par comparaison aux verres), puis dans le cas de systèmes plus complexes comme les ternaires $SiO_2$-PbO-$R_2O$ (R : alcalin) et $SiO_2$-PbO-$Al_2O_3$. L'effet de ces évolutions structurales sur certaines propriétés du verre ou du liquide sera également abordé, à l'exception des propriétés optiques (transmission, coloration) qui sont présentées dans le chapitre 4.

## 3.2. L'élément plomb et les oxydes de plomb

Nous présentons ici des éléments clés permettant de bien comprendre la chimie du plomb dans les oxydes aussi bien cristallisés que vitreux, en soulignant particulièrement l'importance et les conséquences (propriétés physiques et structurales) de l'existence de la paire électronique $6s^2$ libre de l'ion $Pb^{2+}$.

### 3.2.1. *Le plomb élémentaire et ses propriétés particulières*

#### 3.2.1.1. *Les degrés d'oxydation du plomb et l'ion $Pb^{2+}$*

Le plomb (numéro atomique 82) est l'élément stable le plus lourd de la colonne 14 du tableau périodique. De par sa configuration électronique ([Xe] $4f^{14}\, 5d^{10}\, 6s^2\, 6p^2$), il présente trois degrés d'oxydation communs : $Pb^0$, $Pb^{2+}$ et $Pb^{4+}$ dont les domaines de stabilité peuvent être évalués par les potentiels standards des réactions d'ionisation en solution :

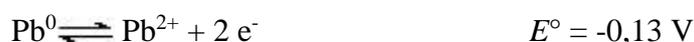

$$Pb^0 \rightleftharpoons Pb^{2+} + 2\,e^- \qquad E° = -0{,}13\ V$$

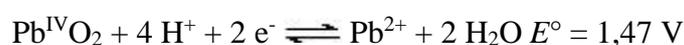

$$Pb^{IV}O_2 + 4\,H^+ + 2\,e^- \rightleftharpoons Pb^{2+} + 2\,H_2O \qquad E° = 1{,}47\ V$$

Selon ces deux équations, le plomb bivalent est donc stable dans un large domaine de potentiel, qui englobe notamment celui du couple $O_2/O^{2-}$ (1,12 V). Il se rencontre donc à ce degré d'oxydation dans la grande majorité des minéraux plombifères comme la litharge, le massicot (PbO) et la galène (PbS) qui en est le principal minerai, ainsi que dans divers carbonates, silicates, phosphates ou sulfates et dans les verres d'oxydes (silicates, borates). Le plomb tétravalent, beaucoup plus rare, se trouve dans le minium ($Pb_3O_4$, ou $Pb^{IV} Pb^{II}_2 O_4$) et la plattnérite ($PbO_2$), surtout connue comme le matériau d'électrode positive des batteries plomb - acide sulfurique. Plus rare encore est le plomb métallique (ou natif). C'est donc logiquement qu'à l'instar de la nature, les arts du feu pratiqués sous un potentiel oxydant proche de celui de l'atmosphère permettent d'élaborer des solides comportant exclusivement des ions $Pb^{2+}$ comme c'est le cas pour les verres à base d'oxydes.

En coordinence 8, l'ion $Pb^{2+}$ présente un rayon de 1,43 Å, comparable à celui du strontium $Sr^{2+}$ (1,40 Å) (Shannon 1976). Comme pour les ions alcalino-terreux lourds $Ca^{2+}$, $Sr^{2+}$ et $Ba^{2+}$ - volumineux et peu chargés - ses oxydes ont un comportement basique qui les rend réactifs vis-à-vis des oxydes acides ($B_2O_3$, $CO_2$, $SiO_2$, $P_2O_5$, $SO_3$, ...) ou amphotères ($H_2O$, $Al_2O_3$, $Fe_2O_3$, ...) et donne lieu à la formation de nombreux oxydes mixtes. Les verres à base d'oxyde de plomb, bien que n'obéissant pas à des compositions définies, s'inscrivent dans ce contexte. Cependant, la configuration électronique externe de l'ion $Pb^{2+}$ ([Xe] $4f^{14}$ $5d^{10}$ $6s^2$) diffère radicalement de celle, saturée, des ions de la 2$^{ème}$ colonne (alcalino-terreux). On peut la schématiser comme une sphère comprenant les orbitales $4f$ et $5d$ et dont le rayon correspond à celui de l'ion $Pb^{4+}$ (1,08 Å) (Shannon 1976), le volume résiduel externe (soit près de 60 %) étant occupé par la seule paire $6s^2$. Fortement écrantés par les orbitales internes, ces deux électrons $6s$ sont à la fois labiles et très sensibles au champ électrique local.

### 3.2.1.2. *La polarisabilité de l'ion $Pb^{2+}$*

La polarisabilité ($\alpha$), qui mesure la capacité d'une particule à se polariser sous l'effet d'un champ électrique $E$ ($P = \alpha.E$, $P$ étant la polarisation induite) est généralement proportionnelle au volume pour des ions de même charge appartenant à une même colonne. Comme le montre la figure 3.2, cette relation est bien respectée dans la série des alcalino-terreux, mais l'ion $Pb^{2+}$ s'en démarque très nettement du fait de la contribution de la paire $6s^2$. Cette polarisabilité élevée s'avère être un paramètre-clé des propriétés physiques, chimiques et structurales de l'ion $Pb^{2+}$, et plus généralement des ions des éléments du bloc *p* présentant une configuration électronique externe $ns^2np^0$ comme $Tl^+$, $Bi^{3+}$ et $Te^{4+}$, isoélectroniques de $Pb^{2+}$.

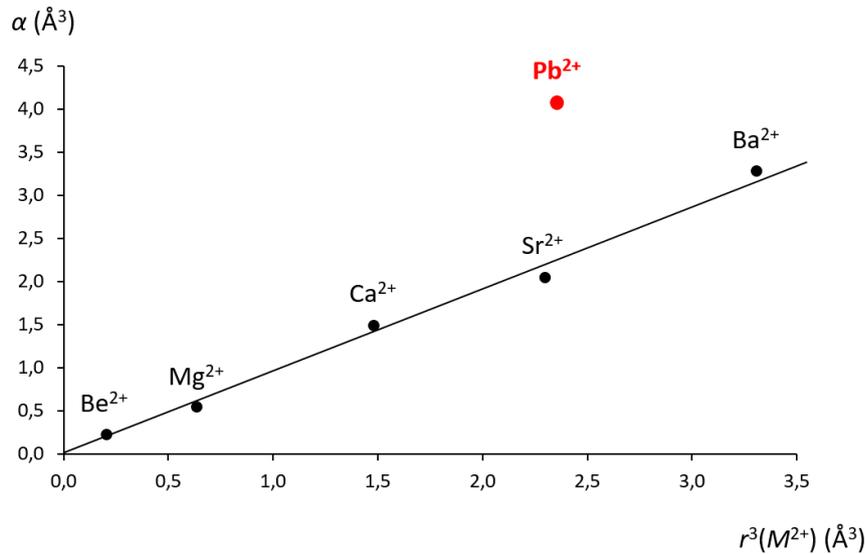

**Figure 3.2.** *Polarisabilité α de l'ion Pb$^{2+}$ et des ions alcalino-terreux (Shannon et Fischer 2016) représentée en fonction du cube du rayon en coordinence 6.*

La polarisabilité détermine notamment la permittivité électrique relative $\varepsilon_r$ comme le montre la relation de Clausius-Mosotti :

$$\frac{\varepsilon_r - 1}{\varepsilon_r + 2} = \frac{N\alpha}{3\varepsilon_0}$$

soit par inversion :

$$\varepsilon_r = \frac{3\varepsilon_0 + 2N\alpha}{3\varepsilon_0 - N\alpha}$$

$N$ étant le nombre de particules par élément de volume et $\varepsilon_0$ la permittivité électrique du vide.

Une des principales applications technologiques des céramiques à base d'oxyde de plomb se trouve logiquement dans l'industrie des composants di-, piézo-, pyro- et ferroélectriques, dont les propriétés sont directement liées à la polarisabilité et où Pb$^{2+}$ a progressivement remplacé Ba$^{2+}$ dans leur composition. Dans le domaine des fréquences optiques, la polarisabilité s'exprime à travers l'indice de réfraction $n$ (= $\sqrt{\varepsilon_r}$). L'effet éminemment positif du plomb II peut être illustré par la série des composés cristallins $M^{II}CO_3$ à structure aragonite (orthorhombique, biaxe), dont l'indice de réfraction moyen, quasi identique (1,62) pour $M$ = Ca, Sr et Ba, s'élève à 1,98 pour la cérusite PbCO$_3$. L'éclat et le pouvoir couvrant d'un matériau pulvérulent étant directement liés à son indice de réfraction, on comprend aisément pourquoi le blanc de plomb (mélange de cérusite et d'hydrocérusite Pb$_3$(CO$_3$)$_2$(OH)$_2$) a été considéré à travers les âges comme le meilleur de tous les pigments blancs (Welcomme *et al.* 2007 ; Stols-Witlox 2011).

De même, l'introduction du plomb dans la composition des verres et des glaçures a notamment pour effet d'en accroître l'indice de réfraction (Chapitre 1, figure 1.3).

### 3.2.2. *Stéréochimie du plomb dans les oxydes*

La théorie de la Répulsion des Paires Electroniques de la Couche de Valence (RPECV)[2] permet notamment de comprendre les distorsions induites par les doublets non liants des éléments légers du bloc *p* et d'expliquer la géométrie de molécules distordues comme l'eau et l'ammoniac. Un phénomène similaire s'observe dans les complexes et les édifices cristallins contenant des ions lourds à configuration $ns^2np^0$, dont les deux formes cristallines communes de PbO (litharge et massicot) sont des exemples typiques.

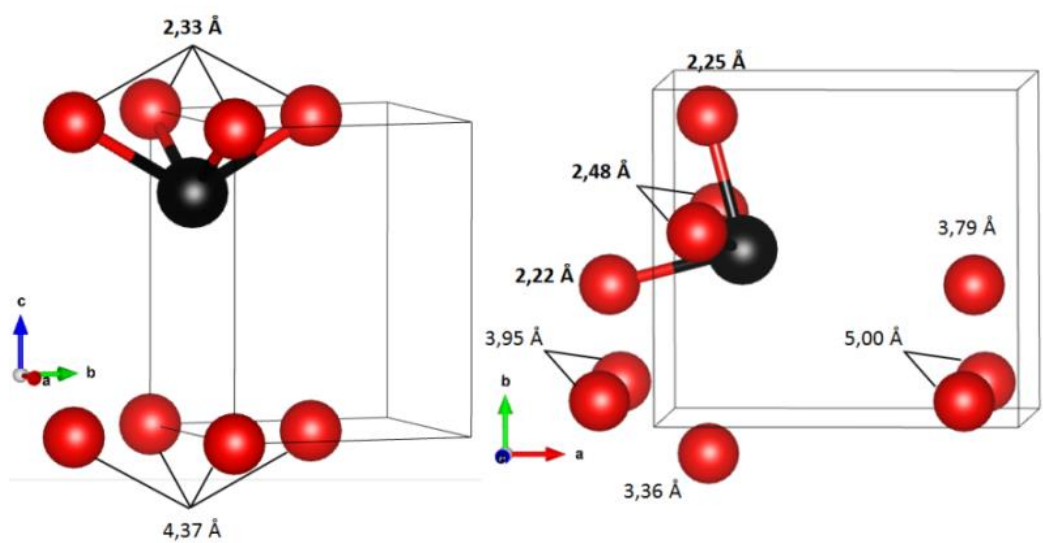

**Figure 3.3.** *Environnement oxygéné de l'ion $Pb^{2+}$ et distances Pb-O dans les deux variétés de PbO : litharge (gauche) et massicot (droite). Pb (en noir) et O (en rouge).*

Dans la litharge (symétrie tétragonale, *P4/nmm*, *a* = 4,004 Å et *c* = 5,071 Å), phase stable à température ambiante, l'ion $Pb^{2+}$ est entouré de huit anions $O^{2-}$ comme dans une structure de type CsCl. Mais ici, au lieu d'un cube, le polyèdre prend la forme d'un pavé très allongé à base carrée, dans lequel le plomb occupe une position fortement excentrée (figure 3.3). Les distances Pb-O, extrêmement disparates (4 × 2,33 Å et 4 × 4,37 Å), traduisent un déséquilibre de répartition des forces de liaison (Brown et Altermatt 1985), orientées à près de 99 % vers les quatre anions les plus proches. La coordinence réelle de $Pb^{2+}$ n'est donc ici que de 4, avec une répartition hémisphérique (dite en "parapluie") propre aux cations à configuration $ns^2np^0$, le plomb occupant ainsi le sommet d'une pyramide à base carrée. Les liaisons courtes présentent ici la même longueur qu'une liaison $Pb^{4+}$-$O^{2-}$ (2,33 Å), ce qui montre que les quatre atomes d'oxygène proches sont en contact avec la sphère formée par les orbitales 4*f* et 5*d* du plomb. Dans le massicot (figure 3.3), forme stable de PbO au-dessus de 490°C, métastable à température ambiante, le plomb adopte également une coordinence 4 hémisphérique, bien que moins régulière du fait de la symétrie orthorhombique (*Pbcm*, *a* = 5,893 Å, *b* = 5,490 Å et *c* = 4,753 Å).

---

[2] En anglais, VSEPR : Valence Shell Electron Pair Repulsion

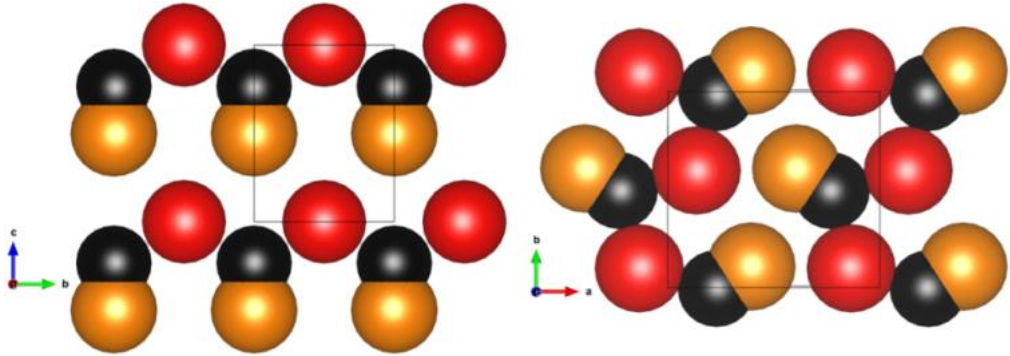

**Figure 3.4.** *Représentation en proportions réelles des ions plomb (en noir) et de la paire 6s² (en orange) dans les plans atomiques de la litharge (gauche) et du massicot (droite). Remarquer l'édifice quasi hexagonal réalisé par les anions O²⁻ (en rouge) et les paires 6s².*

Dans les deux cas, la paire 6$s^2$ se trouve donc repoussée à l'opposé des liaisons courtes en raison de la forte polarisabilité de l'ion $Pb^{2+}$, dans la partie apparemment inoccupée du polyèdre oxygéné (« apparemment » du fait que ces deux électrons, qui constituent une source de diffusion beaucoup moins chargée et moins cohérente que les 78 électrons des orbitales de cœur du plomb regroupés dans un faible volume sont difficilement observables par diffraction X, et encore moins par diffraction neutronique, qui ne « voit » que les noyaux). Du point de vue électrostatique, la paire 6$s^2$ joue cependant un rôle majeur, semblable à celui d'une paire non liante en RPECV, car sa charge négative repousse fortement les anions situés à l'opposé des liaisons fortes, d'où la forte augmentation de la distance Pb-O (4,37 Å) de ce côté du polyèdre (figure 3.3). Le moment dipolaire de l'ion $Pb^{2+}$ affecte ainsi la géométrie de son environnement. La représentation des ions en proportions réelles dans les deux variétés de PbO (figure 3.4) montre qu'elle se trouve excentrée d'environ 1 Å par rapport au noyau et qu'elle occupe sensiblement le même volume qu'un anion. Sa présence dans les volumes apparemment vides des structures permet en réalité, avec les anions oxygène, de réaliser des agencements pseudo hexagonaux proches de la compacité idéale qui contribuent à la stabilité de ces oxydes.

Intéressons-nous maintenant à la façon dont sont connectés les polyèdres PbO$_4$ au sein des deux variétés cristallines de PbO. Dans le cas de la litharge, ces polyèdres sont liés entre eux par les arêtes et arrangés tête-bêche pour minimiser les répulsions électroniques entre les doublets non liants des atomes de plomb (figure 3.5). Dans le massicot, les pyramides sont liées entre elles par les arêtes, et là encore l'arrangement permet de minimiser les répulsions électroniques entre les doublets non liants des ions Pb$^{2+}$. La litharge est la forme la plus stable, car stabilisée par des interactions entre les doublets non liants, orientés de façon antiparallèle entre proches voisins d'un même plan (001). Cette orientation est perdue en partie dans le massicot à cause de la distorsion des polyèdres, ce qui conduit globalement à l'affaiblissement de cette interaction à l'échelle du cristal ; le massicot est donc moins stable.

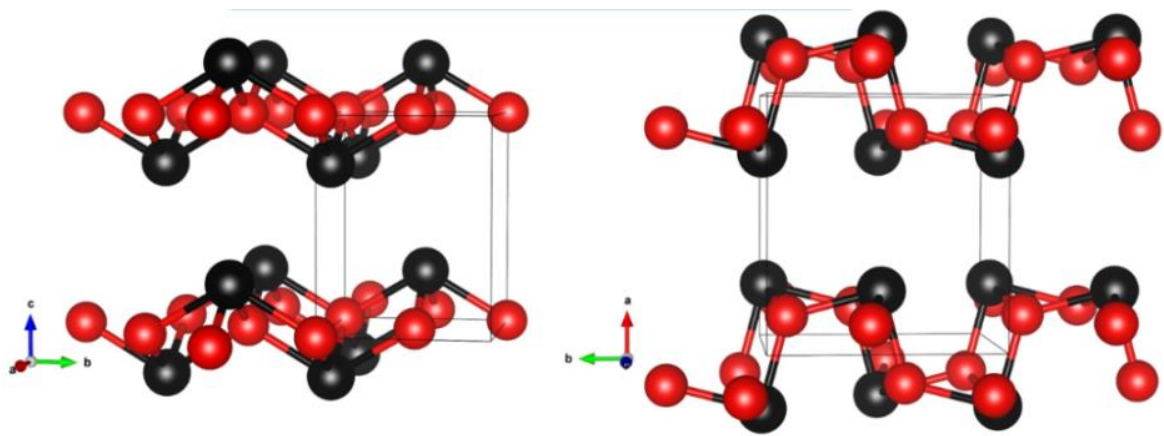

**Figure 3.5.** *Structures cristallines des deux polymorphes de PbO : la litharge (gauche) et le massicot (droite). Pb (en noir) et O (en rouge).*

L'origine du déplacement de l'orbitale 6*s* - normalement sphérique et centrée sur le noyau - dans les ions à configuration $ns^2np^0$ a suscité de nombreux travaux expérimentaux et théoriques depuis ceux de Gillespie et Nyholm dans les années 1950 (Gillespie et Nyholm 1957). Le modèle communément répandu est celui d'une simple hybridation des orbitales 6*s* et 6*p* de Pb$^{2+}$ qui permet au doublet de prendre la forme d'un lobe protubérant, toutefois des études récentes tendent à démontrer l'implication des niveaux 2*p* des anions O$^{2-}$ voisins (Watson *et al.* 1999 ; Waghmare *et al.* 2003). Quel qu'en soit le mécanisme, la distorsion structurale d'un solide cristallin par un effet stéréochimique de « paire libre » obéit *in fine* à une loi simple : minimiser l'énergie réticulaire. Dans la litharge par exemple, le déplacement de la paire 6$s^2$ « met à nu » les orbitales 4*f* et 5*d* à l'opposé, permettant de former quatre liaisons Pb$^{4+}$-O$^{2-}$ courtes, fortement covalentes et directionnelles, globalement plus stables que les huit liaisons à fort taux d'ionicité qui se formeraient dans un édifice symétrique de type CsCl. Si l'on ajoute que son électronégativité (1,87 sur l'échelle de Pauling) est nettement supérieure à celles des éléments alcalins et alcalino-terreux ($\chi_{Na}$ = 0,93 ; $\chi_K$ = 0,82 ; $\chi_{Mg}$ = 1,31 ; $\chi_{Ca}$ = 1,00), on comprendra que le plomb jouera dans les verres un rôle structural beaucoup plus proche de celui des éléments formateurs ou intermédiaires ($\chi_{Si}$ = 1,9 ; $\chi_{Al}$ = 1,61), bien qu'il se rapproche davantage des premiers par sa taille. Comme on le verra plus loin dans ce chapitre (section 3.3.3), dans les verres d'oxydes tels que les silicates, la paire libre des ions $Pb^{2+}$ contribue également à leur incorporation sous forme de polyèdres de faible coordinence (3 ou 4) avec les anions oxygène distribués de façon anisotrope autour du plomb (Fayon *et al.* 1999 ; Rybicki *et al.* 2001 ; Alderman *et al.* 2013).

La même contingence énergétique s'applique cependant parfois à l'encontre de l'activité stéréochimique, lorsqu'une forme structurale stable adaptée à la composition permet d'héberger le cation dans un environnement régulier. La série des carbonates $MCO_3$ à structure aragonite déjà mentionnée plus haut en offre un exemple. Métastable pour $Ca^{2+}$ qui cristallise plus souvent sous forme de calcite, elle est en fait parfaitement adaptée à des cations plus volumineux comme $Sr^{2+}$ et $Ba^{2+}$ pour lesquels elle constitue la forme stable. Or, la structure de la cérusite $PbCO_3$ (Chevrier *et al.* 1992) est quasiment identique à celle de $SrCO_3$ (Pannhorst et Lohn 1970), avec un environnement de $Pb^{2+}$ dont la régularité (2,65 < Pb-O < 2,76 Å) témoigne de l'absence d'activité stéréochimique de la paire 6$s^2$. Cependant, l'absence de polarisation permanente de l'ion $Pb^{2+}$ ne contrarie en rien la forte polarisabilité évoquée préalablement. Notons par ailleurs que l'activité stéréochimique de la paire libre de l'ion $Pb^{2+}$ dépend également de la nature des anions voisins. Ainsi, contrairement à PbO, dans PbS la paire libre des ions $Pb^{2+}$ est stéréochimiquement inactive ce qui serait dû à des interactions orbitalaires anions-cations différentes pour les deux phases (Walsh et Watson 2005). L'activité stéréochimique de la paire 6$s^2$ n'est donc pas systématique dans les structures cristallines contenant du plomb II, mais elle est tout de même fréquente (c'est le cas des silicates de plomb cristallisés comme on le verra dans la Section 3.2), bien que parfois moins prononcée que dans les deux variétés de PbO. $Pb^{2+}$ présente à ce titre un comportement intermédiaire entre ceux de ses voisins isoélectroniques, l'activité stéréochimique étant plus rare et moins marquée chez $Tl^+$ et quasi systématique et forte chez $Bi^{3+}$, sans doute pour des questions de taille et de charge.

Dans le domaine biologique, l'effet de la paire libre contribue largement à la toxicité du plomb, lorsqu'il remplace des cations indispensables au fonctionnement métabolique comme $Ca^{2+}$, $Zn^{2+}$ ou $Fe^{2+}$ au sein des structures des enzymes et des complexes. Cette substitution est d'une part aisée et difficilement réversible car l'activité stéréochimique de la paire 6$s^2$ permet la formation de liaisons fortes avec les ligands, délétère d'autre part car elle en modifie la symétrie au point d'en perturber le fonctionnement (Gourlaouen et Parisel 2007), ce qui rend l'ingestion de plomb particulièrement toxique (saturnisme, chapitres 13 et 14) (Patrick 2006).

## 3.3. Les phases cristallines et les verres du système $SiO_2$-PbO

Dans les verres silicatés à base d'oxyde de plomb utilisés aussi bien dans l'industrie que dans l'artisanat et l'art (verres massifs, glaçures, émaux), $SiO_2$ et PbO représentent les oxydes majoritaires formant la structure du verre. Ainsi dans le verre de cristal, la somme $SiO_2$ + PbO représente généralement de l'ordre de 85% de sa masse (le reste étant principalement constitué d'oxydes alcalins $K_2O$ et $Na_2O$) tandis que dans les glaçures plombifères transparentes déposées en surface des céramiques, cette somme atteint couramment plus de 90% en masse (le reste étant alors principalement constitué d'alumine la plupart du temps (Tite *et al.* 1998 ; chapitre 10). Bien que les verres binaires du système $SiO_2$-PbO n'aient en réalité pas d'utilisation pratique car on rajoute presque toujours à leur composition d'autres oxydes formateurs, modificateurs ou intermédiaires (oxyde de bore, oxydes alcalins, alumine…) en fonction des applications visées (Rabinovich 1976), il apparaît particulièrement important de bien comprendre la structure de ces verres simples en fonction de leur teneur en PbO et en particulier la façon dont les ions $Pb^{2+}$ s'insèrent dans la structure du réseau vitreux silicaté aussi bien pour les faibles teneurs que pour les plus élevées en PbO, si l'on souhaite appréhender la structure de verres silicatés plombifères plus complexes. Cela explique le grand nombre de publications portant sur l'étude structurale des verres du système $SiO_2$-PbO depuis près de 85 ans (Bair 1936) ! Malgré les différentes méthodes d'investigation et de simulation de plus en plus poussées mises en œuvre afin de sonder l'environnement du plomb, du silicium et de l'oxygène dans ces verres simples, de nombreuses interrogations demeurent sur la façon dont les ions $Pb^{2+}$ sont incorporés dans ces verres et particulièrement pour les teneurs les plus élevées en PbO comme on va le voir ici. Mais avant de nous intéresser au système vitreux (c'est-à-dire au liquide surfondu figé lors de la trempe et qui est hors équilibre), penchons-nous d'abord sur le système binaire à l'équilibre thermodynamique en fonction de la température et de sa composition, et également sur la structure des phases cristallines mixtes existant dans ce système et plus particulièrement sur le mode d'incorporation des ions $Pb^{2+}$ dans leur réseau cristallin.

### 3.3.1. *Le système binaire $SiO_2$-PbO à l'équilibre*

Etant donné l'importance de ce système, des diagrammes de phases ont été proposés pour le binaire $SiO_2$-PbO dès les années 1930 (Geller *et al.* 1934) puis complétés et affinés par la suite à l'aide de nouveaux résultats expérimentaux et de calculs. Le plus récent est à notre connaissance celui de Shevchenko et Jak (Shevchenko et Jak 2018), couplant calculs et résultats expérimentaux (figure 3.6).

Comme indiqué dans le chapitre 1, ce diagramme montre le très net effet fondant de PbO lors de son ajout à SiO$_2$ (forte décroissance de la courbe verte correspondant au liquidus du système). On peut remarquer cependant que pour des teneurs en PbO comprises entre 40 et 80 mol%, la température de liquidus (T$_{liquidus}$) évolue peu et reste comprise entre 730 et 750°C mais elle réaugmente un peu quand on se rapproche de la phase PbO pure (massicot, T$_{fusion}$ = 886°C).

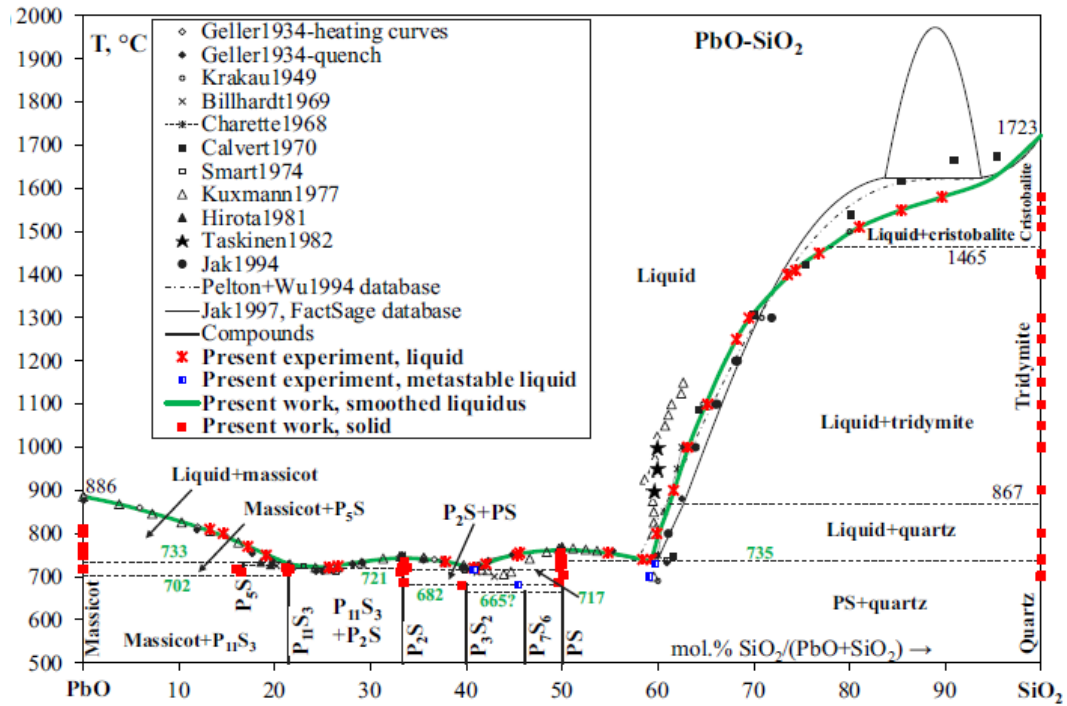

**Figure 3.6.** *Diagramme de phase du système PbO-SiO$_2$ (mol%) au-dessus de 500°C (Shevchenko et Jak 2018) couplant plusieurs sources expérimentales et calculées de la littérature avec les résultats expérimentaux récents (les points rouges et bleus correspondent aux compositions et températures explorées par les auteurs de l'étude et la courbe de liquidus en vert est le résultat d'un lissage réalisé dans cette même étude). Dans ce diagramme est également reporté en trait plein le diagramme de phase calculé par la même équipe dans une précédente étude (Jak et al. 1997). Massicot : phase PbO stable à haute température, pour les phases cristallines mixtes : P = PbO et S = SiO$_2$.*

Tant que la teneur en PbO reste inférieure à 50 mol%, les seules phases cristallines stables susceptibles de se former au sein du système sont les différentes variétés cristallines de la silice (cristobalite, tridymite, quartz) et le métasilicate de plomb PbSiO$_3$ (alamosite) à fusion congruente. Il n'existe donc aucune phase cristalline mixte stable renfermant moins de 50 mol% de PbO. Notons que contrairement aux prévisions des calculs de Jak et al. (Jak *et al.* 1997) (courbe noire, figure 3.6), la courbe expérimentale de liquidus ne montre pas de dôme de démixtion stable indiquant l'existence de phénomènes de séparation de phase au sein du liquide ($T > T_{liquidus}$). Toutefois, la forme particulière en S de la courbe de liquidus (courbe verte, figure 3.6) laisse penser qu'une démixtion métastable peut intervenir au sein du liquide surfondu ($T < T_{liquidus}$) (Calvert et Shaw 1970) comme dans le système $SiO_2$-$BaO$ (Seward *et al.* 1968). Le diagramme montre que plusieurs phases cristallines mixtes sont susceptibles de se former dans ce système en plus de $PbSiO_3$ pour des teneurs en PbO supérieures à 50 mol% :

- A haute température, près de la courbe de liquidus : $Pb_2SiO_4$ (cyclosilicate de plomb, à fusion congruente), $Pb_{11}Si_3O_{17}$ (à fusion congruente) et $Pb_5SiO_7$ (à fusion non congruente). La phase $Pb_4SiO_6$ reportée dans les diagrammes de phase plus anciens (Smart et Gasser 1974) ne serait pas stable (Hirota et Hasegawa 1981).

- A plus basse température, dans le domaine du solidus : $Pb_7Si_6O_{19}$ et $Pb_3Si_2O_7$ (ces deux phases se décomposent avant de fondre).

### 3.3.2. *Les phases cristallines du système binaire SiO$_2$-PbO*

Nous présenterons ici des éléments sur l'évolution de la structure des phases cristallines mixtes mentionnées dans le paragraphe précédent en fonction de la teneur en oxyde de plomb en nous basant sur les déterminations structurales reportées dans la littérature[3] (mol% PbO): l'*alamosite* $PbSiO_3$ (50%) (Boucher et Peacor 1968), $Pb_7Si_6O_{19}$ (53,8%) (Siidra *et al.* 2014), $Pb_3Si_2O_7$ (60%) (Petter et Harnik 1971), $Pb_2SiO_4$ (66,6%) (Dent Glasser *et al.* 1981), $Pb_{11}Si_3O_{17}$ (78,6%) (Hirota et Hasegawa 1981 ; Kato 1982). Trois de ces structures sont représentées en figure 3.7. $Pb_5SiO_7$ (83,3%), stable uniquement à haute température (Hirota et Hasegawa 1981), est la seule dont la structure reste non résolue. Nous nous focaliserons en particulier sur l'effet de l'ajout de PbO sur le réseau silicaté et sur la façon dont les ions $Pb^{2+}$ s'incorporent dans cette structure (type de sites occupés, existence de connexion entre eux et avec les unités silicatées présentes). L'absence d'oxydes mixtes cristallisés avec une teneur inférieure à 50 mol% PbO (soit 79 masse%) est une tendance qui se retrouve parmi les silicates complexes connus (exception : la margarosanite $PbCa_2Si_3O_9$ présente un taux de 43 mol% PbO). Nous ne pourrons donc estimer le comportement structural des verres de type cristal et des glaçures au plomb (renfermant respectivement des teneurs en PbO de l'ordre de 10 et 30 mol%) à partir des phases cristallines que par extrapolation, ou interpolation avec le pôle pur $SiO_2$. Cependant comme on le verra dans la section 3.3.3.1, des verres binaires du système $SiO_2$-$PbO$ renfermant jusqu'à 83 mol% PbO peuvent être obtenus, c'est-à-dire pour des teneurs aussi élevées en PbO que les phases cristallines (figure 3.7). L'examen des structures cristallines montre que la paire $6s^2$ de $Pb^{2+}$est systématiquement active d'un point de vue stéréochimique, ainsi le nombre moyen de liaisons courtes ($d_{Pb-O}$< 2,5 Å, soit avec une valence de liaison supérieure à 0,35 u.v.[4]) ne varie que de 2,3 à 3,0 par ion $Pb^{2+}$, sans qu'aucune corrélation avec le taux de PbO n'apparaisse.

---

[3] Bien que la majorité des études structurales menées sur les phases cristallines binaires du système $SiO_2$-PbO aient été réalisées par diffraction des rayons X, des études RMN menées sur les noyaux $^{29}Si$ et $^{207}Pb$ ont permis de corréler les changements de déplacements chimiques à l'évolution de l'environnement du silicium et du plomb au sein de la structure des cristaux fournissant ainsi des données très utiles pour comprendre l'évolution de l'environnement de ces mêmes espèces au sein des verres (Bessada *et al.* 1994, Fayon *et al.* 1997).

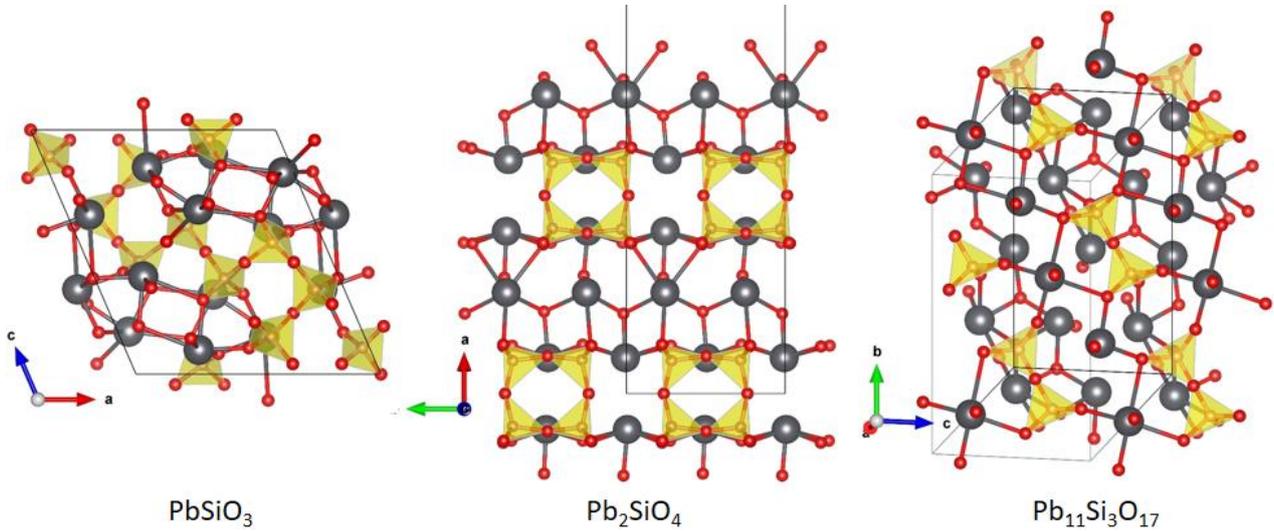

**Figure 3.7.** *Structures cristallines de trois silicates de plomb (1/2 maille est représentée en profondeur) renfermant des teneurs croissantes en PbO de gauche à droite (50; 66,6 et 78,6 mol%) montrant les divers modes de connexion entre tétraèdres $SiO_4$ (en jaune) ainsi que l'asymétrie des polyèdres de coordination $PbO_n$. Pb (en noir) et O (en rouge).*

En dispersant les tétraèdres $SiO_4$, l'ajout de PbO est supposé réduire mécaniquement leur réticulation. Ainsi le réseau silicaté, formé de chaines *méta* infinies dans $PbSiO_3$ évolue vers des unités *di*silicates dans $Pb_3Si_2O_7$, selon un schéma identique à celui observé par ajout de CaO. Pour ces deux familles, le taux d'atomes d'oxygène pontants (impliqués dans des liaisons Si-O-Si) évolue conformément au modèle théorique selon une loi en $(2 - 3x) / (2 - x)$, avec $x = M^{II} / (M^{II} + Si)$ qui rend compte de la dépolymérisation du réseau silicaté (figure.3.8). Il en est de même pour $Pb_7Si_6O_{19}$, dont la structure plus complexe se développe en $Pb_{21}(Si_7O_{22})_2(Si_4O_{13})$. L'analyse des structures cristallines suggère donc que dans les verres et glaçures au plomb (0,1 <$x$< 0,3), l'oxyde de plomb a un effet dépolymérisant similaire à celui d'un modificateur-type. Pour $x = 2/3$, seuil théorique de la dépolymérisation totale (figure 3.8), les tétraèdres silicatés deviennent effectivement indépendants dans $Ca_2SiO_4$ (*ortho*silicate de calcium), mais son homologue au plomb adopte une structure radicalement différente, de type *oxy-cyclo*silicate ($Pb_2OSiO_3$, figure 3.7) qui permet la persistance de pontages Si-O-Si (cycles constitués de 4 tétraèdres). Il en est de même dans $Pb_{11}Si_3O_{17}$ ($Pb_{11}O_6(Si_2O_7)(SiO_4)$), avec un taux de plomb encore plus élevé qui renferme des dimères silicatés (figure 3.7). Cette singularité s'explique vraisemblablement par la stéréochimie de la paire $6s^2$ qui permet d'établir des liaisons fortes en direction opposée des unités silicatées, certains anions étant ainsi accaparés par trois ou quatre cations $Pb^{2+}$ et isolés du réseau silicaté (figure 3.7), ce dernier peut donc maintenir un certain degré de polymérisation par exemple sous forme de cycles ou de dimères (figure 3.7).

---
[4] u.v. (unité de valence de liaison)

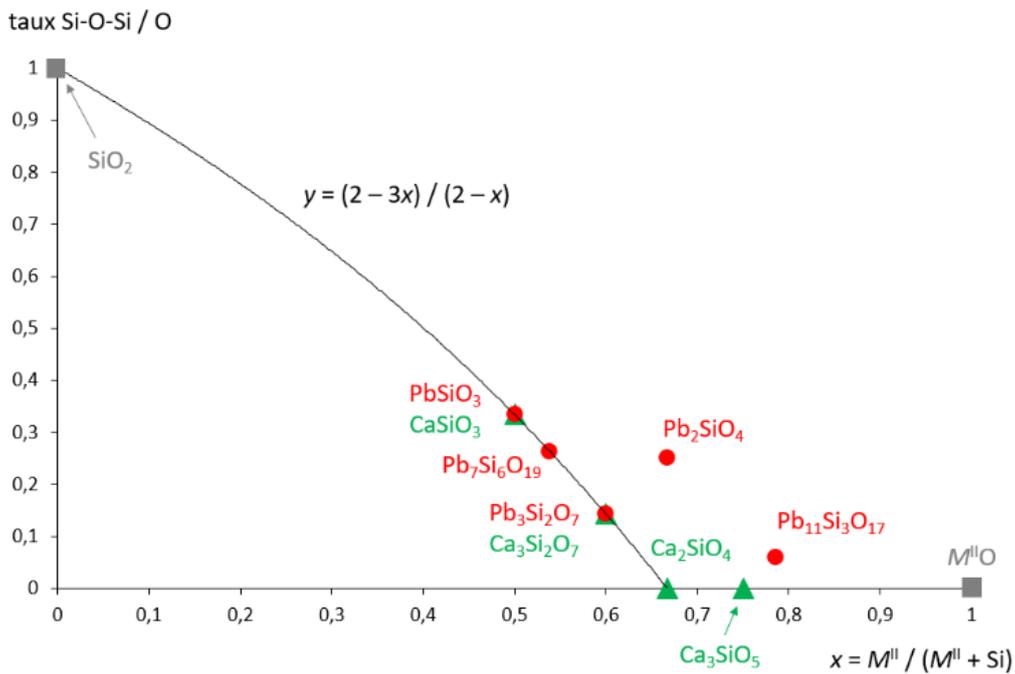

**Figure 3.8.** *Evolution du taux de pontages Si-O-Si par atome d'oxygène en fonction du rapport molaire $M^{II} / (M^{II} + Si)$ pour M = Pb et Ca dans les silicates de plomb (●) ou de calcium (▲) cristallisés.*

Les liaisons fortes Pb-O (avec des ions $O^{2-}$ non connectés au silicium) sont à l'origine d'une seconde forme de réticulation, inexistante avec des modificateurs alcalins ou alcalino-terreux. Elle est cependant plus difficile à définir que celle du réseau silicaté du fait de l'irrégularité de l'environnement du plomb. En ne prenant en compte que les liaisons Pb-O d'une longueur inférieure à 2,5 Å, on observe la formation de pontages Pb-O-Si et Pb-O-Pb, dont les évolutions sont représentées figure 3.9. La « ségrégation » du plomb qui s'opère au-delà de $x = 0,66$ provoque une diminution marquée des premiers, au bénéfice des seconds. Dans le domaine des verres et glaçures (0,1 <$x$< 0,3), l'interpolation entre $SiO_2$ et les silicates de plomb cristallisés les plus proches suggère l'existence d'une réticulation primaire significative, de l'ordre de 0,2 à 0,7 pontages « forts » Pb-O-Si par oxygène. Les pontages Pb-O-Pb seraient en revanche de l'ordre de trois ou quatre fois plus rares (figure 3.9). Au sein des phases cristallines du système $SiO_2$-PbO qui sont toutes très riches en plomb (≥ 50 mol% PbO, figure 3.6) la connexion entre les polyèdres occupés par les ions $Pb^{2+}$ s'effectue généralement par les arêtes (figure 3.7).

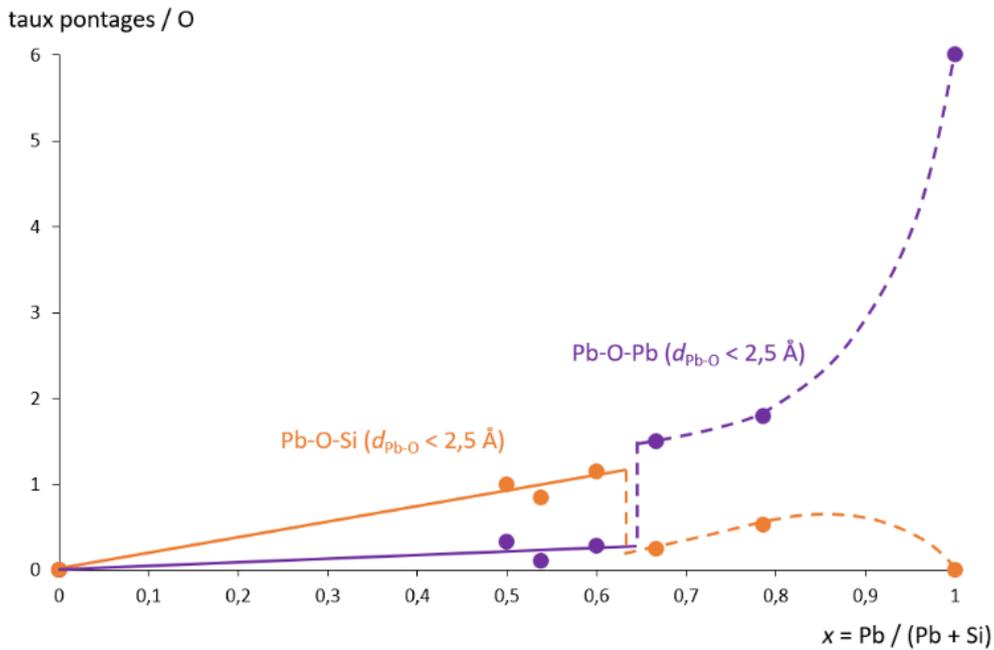

**Figure 3.9.** *Evolution du taux de pontages Pb-O-Si et Pb-O-Pb par atome d'oxygène en fonction du rapport molaire Pb / (Pb + Si) dans les phases cristallines de silicate de plomb.*

Cette approche basée sur la comparaison avec les structures cristallines connues, bien que rendue précaire par leurs taux élevés de PbO (≥ 50 mol%) pour les verresde type cristal et les glaçures plombifères (≤ 30 mol% PbO) mais plus réaliste pour les verres plus riches en PbO renfermant des teneurs dans la même gamme que les phases cristallines (section 3.3.3.1), suggère donc que le plomb contribue à la dépolymérisation du réseau silicaté de la même façon que le calcium, mais que grâce à l'activité stéréochimique de la paire 6$s^2$, il permet la formation de pontages Pb-O-Si basés sur des liaisons Pb-O fortes. Les pontages Pb-O-Pb forts seraient marginaux pour les verres de type cristal et les glaçures plombifères mais auraient une contribution importante et croissante pour les verres renfermant des teneurs en PbO bien plus élevées (figure 3.9).

### 3.3.3. *Les verres du système binaire SiO$_2$-PbO*

#### 3.3.3.1. *Généralités sur les verres du système binaire SiO$_2$-PbO*

Après les phases cristallines de silicate de plomb, intéressons-nous maintenant à l'impact de l'ajout de quantités croissantes de PbO à l'oxyde formateur bien connu qu'est la silice vitreuse SiO$_2$ et plus particulièrement sur sa structure et la façon dont s'incorporent les ions Pb$^{2+}$ dans celle-ci[5] ainsi qu'à l'évolution de certaines propriétés physiques. L'étude de la structure de ces verres et des relations structure-propriétés est particulièrement importante tant du point de vue fondamental qu'appliqué, SiO$_2$ et PbO représentant les deux oxydes majoritaires dans la plupart des verres silicatés à base d'oxyde de plomb (cristallerie, glaçures et émaux, optique…). De plus, le domaine de composition vitrifiable très étendu dans ce système au regard des autres systèmes binaires tels que les silicates alcalins ou alcalino-terreux suscite toujours beaucoup de questionnements au niveau de leur structure et du rôle du plomb en particulier pour les compositions les plus riches en PbO. En effet, il est possible d'obtenir des verres binaires SiO$_2$-PbO par fusion-trempe jusqu'à des teneurs en PbO de plus de 83 mol%[6] (95 masse%) (Feller *et al.* 2010) alors que pour les systèmes SiO$_2$-R$_2$O (R : alcalin) on obtient de plus en plus difficilement des verres au-dessus de 55 mol% R$_2$O (Imaoka et Yamazaki 1963) en raison de la très forte tendance à la cristallisation des liquides surfondus correspondants lors du refroidissement, associée à la fois à une diminution de leur viscosité et à la très forte dépolymérisation du réseau silicaté[7]. Dans le système plombifère, pour les compositions très riches en PbO, on réussit donc à stabiliser par trempe la structure amorphe du liquide bien que SiO$_2$ soit minoritaire. Cela laisse suspecter un comportement structural différent des ions Pb$^{2+}$ en comparaison des ions alcalins et alcalino-terreux au sein de ces verres et en particulier pour les fortes teneurs en PbO (ce qui est en accord avec la comparaison des phases cristallines de silicates alcalino-terreux et de plomb, figure 3.8), bien que, comme eux, son ajout progressif à SiO$_2$ induise à la fois une forte diminution de la température de liquidus du système (figure 3.6) et de la viscosité du liquide (figure 3.10).

---

[5] Comme indiqué plus haut (section 3.2.1.1), le plomb est très majoritairement présent sous forme d'ions Pb$^{2+}$ dans les verres d'oxydes. La présence d'ions Pb$^{4+}$ interviendrait seulement pour des compositions de verres très riches en plomb fondues sous atmosphère très oxydante. Cela va dans le sens de l'évolution de la stabilité des oxydes de plomb avec la température (PbO$_2$< Pb$_3$O$_4$<PbO).

[6] Des verres binaires présentant des teneurs en PbO supérieures à 90 mol% ont parfois été reportés dans la littérature (Ellis *et al.* 1979), mais par manque de résultats d'analyse chimique ces données doivent être prises avec beaucoup de précaution du fait de la très forte tendance à l'évaporation de PbO pour les compositions riches en plomb (> 70-80 masse% PbO) (Andersen 1919 ; Volf 1984). De plus, les liquides en fusion très riches en PbO (> 90 masse%) présentent une très forte tendance à la cristallisation (Faivre et Weiss 1963, p. 587).

[7] Pour les verres binaires SiO$_2$-R$_2$O, si on suppose la formation progressive d'entités SiO$_4$Q$^n$ avec un nombre n décroissant d'atomes d'oxygène pontants (figure 3.1b )lors de l'ajout de R$_2$O, il apparaît qu'entre 50 et 60 mol% R$_2$O il n'y aurait que des entités très dépolymérisées Q$^2$ et Q$^1$(modèle binaire). Au-delà de 66,6 mol% R$_2$O, le verre ne serait alors plus constitué que d'entités Q$^0$ (SiO$_4$$^{4-}$) totalement dépolymérisées. Il est donc possible d'obtenir des verres binaires SiO$_2$-PbO bien au-dessus de cette dernière valeur. Pour plus de détails voir la section 3.3.3.2.1.

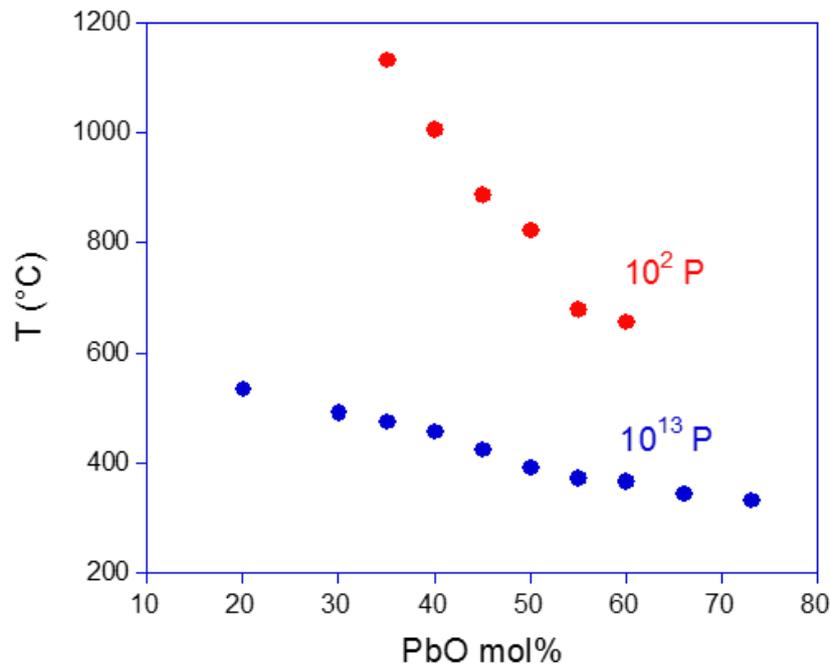

**Figure 3.10.** *Evolution en fonction de la teneur en PbO de la température correspondant aux viscosités de $10^2$ et $10^{13}$ Poises pour les liquides du système $SiO_2$-PbO (d'après les données de Martlew (Martlew 2005)). On constate que la température correspondant à une viscosité donnée décroit avec la teneur en PbO ce qui montre son effet fluidifiant.*

Il est à noter que même si le liquide est toujours homogène pour $T > T_{liquidus}$ (figure 3.6), les compositions riches en $SiO_2$ (> 65 mol%) présentent une tendance à la démixtion (métastable, c'est-à-dire sub-liquidus) au cours du refroidissement du liquide (Calvert et Shaw 1970 ; Vogel 1994). Toutefois, la tendance à la séparation de phase dans ces verres binaires est supprimée lors de l'ajout de $K_2O$ (comme c'est le cas dans le verre de cristal) ce qui peut s'expliquer par la quasi-absence de tendance à la démixtion dans le binaire $SiO_2$-$K_2O$ (Charles 1966) en raison de la faible force de champ[8] du cation $K^+$ (F ~ 0,14 Å$^{-2}$) par rapport au cation $Pb^{2+}$ (F ~ 0,36 Å$^{-2}$) (Vogel 1994). La tendance marquée des verres binaires $SiO_2$-PbO à faible teneur en PbO à la démixtion couplée à leur température de liquidus très élevée (figure 3.6) nécessitant des températures de fusion de l'ordre de 1600-1700°C explique le peu d'études reportées dans la littérature sur les compositions de silicate de plomb très riches en silice (Bair 1936 ; Calvert et Shaw 1970). Un certain nombre d'études ont également été réalisées sur la cristallisation des verres binaires $SiO_2$-PbO avec ou sans agent nucléant (tel que $P_2O_5$) mais ne seront pas développées ici (Carr et Subramanian 1982a ; Carr et Subramanian 1982b ; Lippmaa *et al.*

---

[8] On peut définir la force de champ F d'un cation $M^{z+}$ comme étant le rapport F = $z/d(M-O)^2$, d(M-O) étant la distance moyenne entre M et les anions $O^{2-}$ premiers voisins. Les valeurs de force de champ pour les ions $K^+$ et $Pb^{2+}$ ont été calculées à partir de valeurs moyennes de distance d(M-O) dans des verres de silicates binaires (d(K-O) ~ 2,64Å et d(Pb-O) ~ 2,40Å).

1982 ; Neiman *et al.* 1982 ; Kaur *et al.* 2013 ; Pena *et al.* 2020). Ainsi, selon Kaur *et al.* (Kaur *et al.* 2013), tant que la teneur PbO< 50 mol% les verres résistent bien à la cristallisation au chauffage alors que lorsque la teneur en PbO≥ 50 mol% la cristallisation majoritaire de phases telles que $PbSiO_3$ et $Pb_2SiO_4$ a été mise en évidence. Avant de rentrer plus avant dans les détails structuraux et en particulier sur la façon dont les ions $Pb^{2+}$ s'insèrent dans la structure vitreuse et comment réagit le réseau silicaté (section 3.3.3.2), il est intéressant de présenter comment l'ajout progressif de PbO à la silice affecte certaines propriétés physiques du verre et du liquide (Kaur *et al.* 2013).

En plus de l'effet sur la viscosité du liquide en fusion dont nous avons parlé plus haut (forte diminution avec la teneur en PbO, figure 3.10), mettant en évidence l'effet fluidifiant de l'oxyde de plomb comme c'est le cas pour les oxydes alcalins et l'oxyde de bore, l'incorporation de teneurs croissantes en PbO à la silice conduit à une diminution de la température de transition vitreuse, de la résistivité électrique des verres et de la tension superficielle des liquides associés. Par ailleurs, le volume molaire tend aussi à diminuer jusqu'à 60 mol% PbO puis augmente ensuite légèrement (Shelby 1983 ; Kaur *et al.* 2013 ; Ben Kacem 2017). L'effet de la teneur en plomb sur les propriétés mécaniques des verres binaires $SiO_2$-PbO a également été étudié montrant par exemple une tendance à la décroissance de leur dureté et de leur module d'Young lorsque la concentration en PbO augmente (Yoshihoto et Soga 1987 ; Kaur *et al.* 2013). L'augmentation de la densité et de l'indice de réfraction présentées dans le chapitre 1 s'explique simplement par la masse atomique élevée du plomb par rapport au silicium et à l'oxygène (densité) et par la forte polarisabilité de l'ion $Pb^{2+}$ et sa masse atomique élevée (indice de réfraction). Notons également que le coefficient de dilatation des verres binaires $SiO_2$-PbO croît linéairement avec la teneur molaire en PbO (Greenough *et al.* 1981 ; Shelby 1983) mais reste toujours inférieur à celui des verres binaires $SiO_2$-$Na_2O$ (Scholze 1980), cette propriété étant importante en particulier dans le cas des glaçures (problème de tressaillage). Le cas de la durabilité chimique en milieu aqueux des verres renfermant de l'oxyde de plomb a été brièvement abordé dans le chapitre 1 (section 1.1.8) et est présenté en détails dans le chapitre 13 pour les verres industriels renfermant à la fois du plomb et des ions alcalins (verre de cristal), ce point ne sera donc pas développé ici. Cependant, concernant les verres binaires $SiO_2$-PbO sans alcalins, on peut indiquer l'étude réalisée par El-Shamy et Taki-Eldin (El Shamy et Taki-Eldin 1974) sur la durabilité chimique des verres en fonction du pH qui montre une évolution non monotone de leur tendance à la lixiviation en fonction de leur teneur en PbO. On peut également citer les travaux plus récents de Mizuno et al. (Mizuno *et al.* 2005) sur la durabilité des verres binaires $SiO_2$-PbO réalisés en milieu acide que ces auteurs ont tenté de relier à la distribution des unités structurales PbO$_n$ à base de plomb au sein de la structure vitreuse en fonction de la concentration en plomb.

3.3.3.2. *Structure des verres du système binaire SiO$_2$-PbO*

Selon certains travaux reportés dans la littérature (Sorokina *et al.*1996) le rôle structural du plomb au sein des verres silicatés binaires évoluerait avec la teneur en PbO, passant de modificateur à formateur, ce qui permettrait d'expliquer la plus grande extension du domaine vitreux des silicates de plomb par rapport aux silicates alcalins évoqués plus haut (section 3.3.3.1). Avant de présenter plus précisément ces travaux ainsi que d'autres études qui soutiennent que le rôle structural du plomb n'évoluerait pas fondamentalement en fonction de sa concentration (Smets et Lommen 1982 ; Mizuno *et al.* 2005 ; Ben Kacem *et al.* 2017) et proposent différents modèles structuraux, penchons-nous sur l'oxyde de plomb PbO seul et sur ses capacités ou non à conduire à un verre par fusion-trempe.

Précisons tout d'abord que PbO n'est pas un oxyde formateur car, à la différence de la silice par exemple, il ne conduit pas par fusion + trempe à la formation d'un verre. Cela peut être dû à plusieurs raisons. Tout d'abord, la très faible viscosité de PbO liquide au voisinage de sa température de fusion ($\eta$~0,01Pa.s (0,1P) (Sinn 1979)) supérieure seulement d'un facteur 10 à celle de l'eau à température ambiante ($\eta$~ 0,001Pa.s (0,01P) (Korson *et al.* 1969)) et très inférieure à celle de la silice fondue au voisinage de sa température de fusion ($\eta$~ 10$^6$Pa.s (10$^7$P) (Martlew 2005)) explique la facilité de cristallisation de PbO à partir du liquide en raison de la forte mobilité des espèces et donc la difficulté de maintenir ce dernier à l'état surfondu et par conséquent d'obtenir un verre par trempe. De plus, d'un point de vue structural, d'après les règles de Zachariasen relatives aux oxydes formateurs M$_x$O$_y$ (Zachariasen 1932) ceux-ci doivent être tels que d'une part les polyèdres MO$_n$ sont faiblement coordinés (3 ou 4) - ce qui correspond à de petits cations M$^{+2y/x}$ fortement chargés tels que Si$^{4+}$, B$^{3+}$ et P$^{5+}$ - et connectés les uns aux autres uniquement par les sommets (non par les arêtes ou les faces) de façon à permettre une flexibilité importante dans l'agencement structural, sans que cela ne conduise à une trop forte augmentation de l'énergie libre du système. D'après ces mêmes règles et pour les mêmes raisons, il est stipulé que les anions oxygène ne doivent pas connecter plus de 2 polyèdres MO$_n$. Les ions Pb$^{2+}$ étant en coordinence 4 dans PbO cristallin comme nous l'avons vu plus haut (figures 3.3 et 3.5), les ions O$^{2-}$ doivent également adopter une coordinence 4 pour respecter l'électroneutralité. Cela explique que les unités PbO$_4$ sont connectées par les arêtes et non par les sommets, ce qui est contraire aux règles de Zachariasen. Cette structure relativement contrainte localement paraît donc peu propice à accepter un désordre important au niveau de l'agencement polyédrique sans que cela ne conduise à augmenter significativement son énergie libre[9], ce qui doit donc favoriser du point de vue thermodynamique l'état cristallisé par rapport à l'état vitreux et pourrait permettre d'expliquer structurellement le fait que PbO n'est pas un oxyde formateur. De même, d'après la théorie de Dietzel (Dietzel 1948) reliant force de champ cationique Fet rôle structural des oxydes dans les verres, il apparaît que l'ion $Pb^{2+}$ présente une force de champ d'environ 0,36 $Å^{-2(10)}$, ce qui classerait PbO dans les oxydes modificateurs pour lesquels F est comprise entre 0,1 et 0,4 $Å^{-2}$ (F est comprise entre 1,5 et 2,0 $Å^{-2}$ pour les oxydes formateurs). Toutefois, il est important de rappeler ici qu'en raison de sa très forte polarisabilité (section 3.2.1.2), du point de vue de ses liaisons avec les ions oxygène plus proches voisins, l'ion $Pb^{2+}$ peut être considéré comme un ion $Pb^{4+}$ avec dans ce cas une force de champ plus élevée F = 0,48 $Å^{-2}$ ce qui classerait alors PbO dans les oxydes intermédiaires (F comprise entre 0,5 et 1,0 $Å^{-2}$) avec des liaisons Pb-O assez directionnelles, à caractère ionocovalent et disposées de façon très dissymétrique dans son environnement (figures 3.3 et 3.5). Ce type de liaison Pb-O est donc très différent de celui existant entre les cations alcalins et alcalino-terreux et les anions $O^{2-}$ dans les oxydes cristallisés (liaisons ioniques réparties assez symétriquement autour des cations).

On comprend alors que le mode d'incorporation des ions $Pb^{2+}$ au sein d'un verre silicaté puisse être bien différent[11] de celui des cations modificateurs traditionnels ($Na^+$, $K^+$, $Ca^{2+}$...) qui sont bien moins polarisables[12] et pour lesquels les liaisons avec les anions oxygène sont plus faibles, ioniques et donc non-directionnelles. C'est d'ailleurs en s'appuyant sur la structure cristalline de PbO et en supposant que l'environnement des ions $Pb^{2+}$ dans les verres $SiO_2$-PbO était également très asymétrique avec des liaisons directionnelles assez fortes en raison de leur très forte polarisation par les anions $O^{2-}$ que Fajans et Kreidl (Fajans et Kreild 1948) ont proposé dès les années 40 que ce type particulier d'environnement du plomb était à l'origine de la plus grande stabilité des liquides du système $SiO_2$-PbO vis-à-vis de la cristallisation (et donc de leur plus grande facilité de vitrification) par rapport aux compositions silicatées alcalines pour lesquelles les processus de cristallisation sont facilités en raison de la réorganisation plus aisée des espèces au sein des liquides correspondants (liaisons plus faibles et non asymétriques entre les cations alcalins et les anions $O^{2-}$)[13]. Pourtant l'étude de la structure des verres binaires $SiO_2$-

---

[9] Il est à noter toutefois qu'en raison du caractère bidimensionnel de la structure cristalline de PbO (litharge et massicot, figure 3.6), celle-ci peut probablement accepter plus facilement des contraintes locales dues à du désordre que les structures cristallines tridimensionnelles comme celles des oxydes alcalins et alcalino-terreux dont le rôle d'oxydes modificateurs est bien établi.
[10] Calcul réalisé d'après la structure de PbO cristallin et les rayons ioniques de Shannon (Shannon 1976) ($rPb^{2+}$ = 0,98Å, $rO^{2-}$ = 1,38Å).
[11] En effet, nous avons vu dans la section 3.3.2 que pour les silicates de plomb cristallisés l'environnement des ions $Pb^{2+}$ est également très fortement asymétrique.
[12] On a en effet les polarisabilités suivantes selon Kordes (en $Å^3$): $Na^+$ (0,175) < $Ca^{2+}$ (0,469) < $K^+$ (0,821) < $Pb^{2+}$ (3,623) (Dimitrov et Komatsu 2010).
[13] Il est intéressant de signaler que selon Vogel (Vogel 1994), le fait que le domaine vitrifiable du système $P_2O_5$-PbO soit moins étendu que celui du système $SiO_2$-PbO (66 mol% contre 83 mol%) s'expliquerait par la très forte contre polarisation exercée par les ions $P^{5+}$ sur les anions $O^{2-}$ du réseau phosphaté qui auraient donc un effet polarisant plus faible sur les ions $Pb^{2+}$ que dans le cas des verres silicatés (l'ion $Si^{4+}$ ayant une force de champ plus faible que l'ion $P^{5+}$), conduisant ainsi à des forces de

PbO avait débuté une dizaine d'années plus tôt en 1936 par diffusion des rayons X (Bair 1936), c'est-à-dire seulement un an après les travaux bien connus de Warren et Loring sur l'étude par diffusion des rayons X de la structure des verres de silicate de sodium (Warren et Loring 1935) et 4 ans après la fameuse publication de Zachariasen sur la structure des verres d'oxydes (Zachariasen 1932). Cependant, le modèle proposé par Bair pour les verres de silicate de plomb (Bair 1936), d'après son traitement des données de diffusion des rayons X, était très proche du modèle structural des verres de silicate de sodium de Warren et Loring (réseau continu aléatoire silicaté avec une distribution homogène des ions $Pb^{2+}$ près des atomes d'oxygène non-pontants des tétraèdres $SiO_4$). Depuis ces premiers travaux visant à élucider la structure des verres binaires $SiO_2$-PbO de très nombreuses études ont été réalisées à l'aide de méthodes d'investigation structurale très diverses telles que la diffusion des rayons X et des neutrons, l'absorption des rayons X (EXAFS[14] et XANES[15]), les spectroscopies XPS[16], RMN[17] multinucléaire ($^{207}$Pb, $^{29}$Si, $^{17}$O), Raman et IR, ainsi que la dynamique moléculaire. Ces études ont confirmé l'hypothèse de Fajans et Kreidl d'un environnement très asymétrique des ions $Pb^{2+}$ au sein des verres silicatés, comme par exemple l'étude de Mydlar *et al.* dès 1970 (Mydlar *et al.* 1970). Bien qu'aujourd'hui l'ensemble des auteurs s'accordent à dire qu'au sein des verres de silicate de plomb la majorité des ions $Pb^{2+}$ sont localisés dans des environnements très asymétriques voisins de ceux existant dans les cristaux de PbO et de silicate de plomb (figures 3.3, 3.5 et 3.7) et qu'ils ont donc comme dans ces phases cristallines leur paire $6s^2$ active (sections 3.2.2 et 3.3.2), des divergences demeurent entre les auteurs quant au rôle structural du plomb, à son type d'environnement et à son évolution en fonction de la teneur en PbO dans le verre. Une partie de ces travaux structuraux sur les verres binaires $SiO_2$-PbO a fait l'objet de synthèses bibliographiques anciennes (Rabinovich 1976) ou plus récentes (Alderman 2013). A partir de certaines de ces études et notamment des plus récentes, nous nous pencherons dans la suite de ce chapitre tout d'abord sur la façon dont le réseau vitreux silicaté est affecté par l'ajout de quantités croissantes d'oxyde de plomb, puis nous nous focaliserons sur le rôle structural des ions $Pb^{2+}$ et sur l'évolution de leur environnement et de leur distribution dans le réseau vitreux.

3.3.3.2.1. Le réseau vitreux silicaté

Avant tout ajout d'oxyde de plomb la structure du verre de silice pure est constituée uniquement de tétraèdres $SiO_4$ connectés les uns aux autres par leur sommets formant un réseau continu

---

liaisons Pb-O plus faibles et une coordinence plus élevée pour les ions $Pb^{2+}$ avec les atomes d'oxygène associés au phosphore (Hoppe *et al.* 2003). Sur la structure et les propriétés des verres du système binaire $P_2O_5$-PbO voir par exemple les références (Dayanand *et al.* 1996 ; El-Egili *et al.* 2003 ; Matori *et al.* 2013).
[14] EXAFS = Extended X-ray Absorption Fine Structure
[15] XANES = X-Ray Absorption Near Edge Structure
[16] XPS = X-ray Photoelectron Spectroscopy
[17] RMN = Résonance Magnétique Nucléaire (RMN MAS = RMN avec rotation à l'angle magique ; RMN MQMAS = RMN multi-quanta avec rotation à l'angle magique)

comme on l'a vu au début de ce chapitre (figure 3.1a). Chaque tétraèdre $SiO_4$ présente dans ce cas quatre atomes d'oxygène pontants. Afin d'introduire la nomenclature relative aux différents types de tétraèdres $SiO_4$ présents dans les verres constitués de silice et d'un ou plusieurs autres oxydes comme PbO, considérons le cas bien connu des verres binaires $SiO_2$-$Na_2O$. Lorsqu'on ajoute $Na_2O$ au verre de silice des liaisons Si-O-Si sont rompues ($Na_2O$ a un rôle d'oxyde modificateur) : le réseau vitreux silicaté devient partiellement dépolymérisé et des atomes d'oxygène non-pontants- dont la charge négative est compensée localement par les ions $Na^+$- sont formés (figure 3.1b). Pour rendre compte de la dépolymérisation progressive des tétraèdres $SiO_4$ en fonction de la teneur en $Na_2O$, on introduit la notation $Q^n$ qui représente les tétraèdres $SiO_4$ comptant n atomes d'oxygène pontants. Un verre de silice pure ne contient ainsi que des unités $Q^4$ et pour les verres binaires une unité $Q^3$ correspond à un tétraèdre $SiO_4^-$, une unité $Q^2$ à un tétraèdre $SiO_4^{2-}$, une unité $Q^1$ à un tétraèdre $SiO_4^{3-}$ et une unité $Q^0$ à un tétraèdre $SiO_4^{4-}$ complètement isolé du reste du réseau silicaté. Plus on a d'unités $Q^n$ avec n faible, plus il devient difficile de vitrifier le liquide car il est de plus en plus fluide et s'organise facilement sous forme de cristaux. En utilisant cette notation, on peut écrire l'équation chimique qui correspond à l'introduction de $Na_2O$ dans un verre de silice pure comme: $2Q^4 + Na_2O \rightarrow 2Q^3$. Il s'agit en fait d'une réaction acido-basique (au sens de l'oxo-acidité (Flood et Förland 1947)) entre l'oxyde basique $Na_2O$ donnant des anions $O^{2-}$ et l'oxyde acide $SiO_2$ qui les acceptent. Par ailleurs, il existe au sein du liquide en fusion un équilibre (réaction de disproparnation) entre les unités $Q^n$ ($2Q^n \leftrightarrow Q^{n-1} + Q^{n+1}$) qu'on retrouve dans le verre obtenu après trempe (Maekawa *et al.* 1991). Décrire le réseau silicaté consiste notamment à étudier son degré de polymérisation, c'est-à-dire à suivre de façon qualitative ou quantitative l'évolution des proportions des différentes unités $Q^n$ qui le constituent. Dans le cas des compositions silicatées simples cela peut se faire par exemple au moyen des spectroscopies RMN $^{29}Si$ et Raman (Maekawa *et al.* 1991 ; O'Shaughnessy *et al.* 2020). Tandis que le suivi de l'évolution des proportions des différents types d'atomes d'oxygène (pontants, non-pontants ou non connectés au réseau silicaté) peut être réalisé au moyen des spectroscopies RMN $^{17}O$ (après enrichissement des verres en $^{17}O$)[18] et XPS par exemple (Dalby *et al.* 2007 ; Lee et Kim 2015).

Dans le cas qui nous intéresse, plusieurs études se sont penchées sur l'évolution de la structure du réseau silicaté des verres binaires de silicate de plomb et mettent clairement en évidence le fait que l'introduction de PbO conduit à une dépolymérisation croissante du réseau silicaté. Par exemple, Bessada *et al.* (Bessada *et al.* 1994), Fayon *et al.* (Fayon *et al.* 1 998) et Fayon (Fayon

---

[18] Etant donné la très faible abondance naturelle de l'isotope $^{17}O$ (0,038%) seul détectable par RMN en raison de son spin nucléaire I non nul (I = 5/2) par rapport à l'isotope $^{16}O$ (99,75%) à spin nucléaire nul, il est nécessaire d'enrichir les verres en $^{17}O$. Dans le cas des verres silicatés cela se fait classiquement en préparant de l'oxyde de silicium $SiO_2$ enrichi, par exemple à partir d'un mélange de $SiCl_4$ et d'eau enrichie en $^{17}O$.

1998) ont utilisé la RMN $^{29}$Si pour comparer les spectres des verres à ceux des phases cristallines et pour quantifier les unités $Q^n$ présentes dans les verres renfermant de 31 à 73 mol% PbO (figures 3.11 et 3.12).

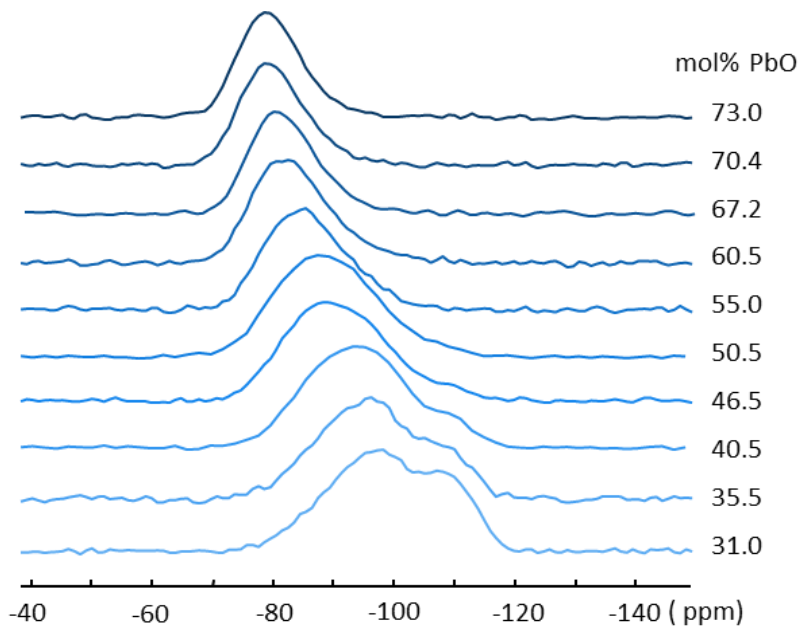

**Figure 3.11.** *Evolution des spectres RMN MAS $^{29}$Si de verres de silicate de plomb en fonction de leur teneur en PbO (Fayon 1998). (Champ magnétique 7 T, fréquence de rotation 10 kHz) (Avec l'aimable autorisation de F. Fayon)*

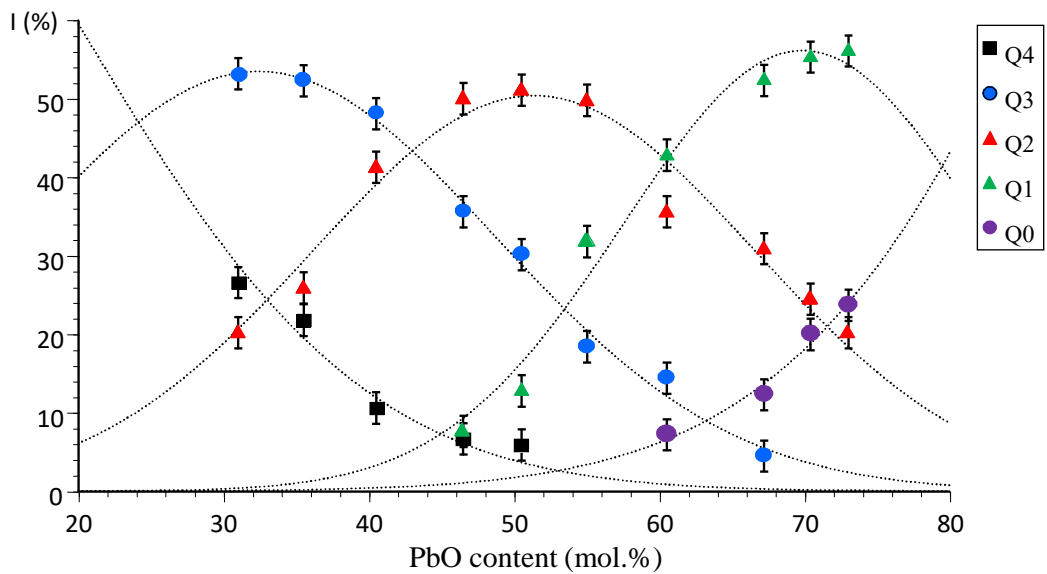

**Figure 3.12.** *Evolution des proportions relatives des unités $Q^n$ déterminées à partir des spectres RMN MAS $^{29}$Si de verres de silicate de plomb (figure 3.11) en fonction de leur teneur en PbO (les courbes en pointillées sont uniquement un guide pour l'œil) (Fayon 1998) (Avec l'aimable autorisation de F. Fayon)*

Il a ainsi été montré que de 31 mol% jusqu'à environ 40 mol% PbO les unités $Q^3$ étaient majoritaires et que la proportion d'unités $Q^2$ augmentait principalement au détriment des unités $Q^4$ (Fayon 1998 ; Fayon *et al.* 1998) (figure 3.12). Puis les unités $Q^1$ apparaissent et leur proportion augmente avec la teneur en PbO principalement au détriment des unités $Q^3$ et deviennent majoritaires au-delà de 60 mol% PbO. Pour finir, les unités $Q^0$ apparaissent et constituent les unités majoritaires avec les unités $Q^1$ au-delà de 72 mol% PbO (figure 3.12). Cette évolution structurale a été confirmée plus récemment sur une gamme plus étendue de teneurs en PbO (33-83 mol%) par Feller *et al.* (Feller *et al.* 2010) qui ont montré que le verre préparé avec 83 mol% PbO ne renfermait quasiment plus que des espèces $Q^1$ (24%), c'est-à-dire des entités constituées de seulement deux tétraèdres $SiO_4$ connectés par un sommet, et des espèces $Q^0$ (72%), c'est-à-dire des tétraèdres $SiO_4$ connectés à aucun autre tétraèdre $SiO_4$, confirmant ainsi pour les compositions binaires $SiO_2$-PbO la possibilité d'obtenir un état vitreux en l'absence de réseau vitreux silicaté étendu. La comparaison des déplacements chimiques $^{29}Si$ des unités $Q^n$ entre des verres binaires et des phases cristallines de même composition a par ailleurs mis en évidence un nombre d'ions $Pb^{2+}$ seconds voisins des atomes de silicium plus important pour les verres (Bessada *et al.* 1994). Il est par ailleurs intéressant de noter que les constantes d'équilibre des réactions $2Q^n \leftrightarrow Q^{n-1} + Q^{n+1}$ (disproportionation) intervenant au sein du liquide surfondu et figées dans le verre lors de la trempe et qui ont été déterminées par RMN $^{29}Si$ à partir des proportions des différentes unités $Q^n$ dans les verres (figure 3.12) sont d'un ordre de grandeur supérieures à celles des verres de silicates alcalins (Fayon *et al.* 1998), ce qui tend à montrer une plus grande tendance des ions $Pb^{2+}$ à clustériser par rapport aux ions alcalins. Le fait que des unités $Q^2$ et $Q^1$ persistent bien au-delà de 66 mol% PbO (limite au-delà de laquelle ne devraient demeurer que des unités $Q^0$ si PbO transférait totalement ces anions $O^{2-}$ au réseau silicaté par réaction acido-basique) suggère qu'une fraction croissante des atomes d'oxygène de la structure ne sont connectés qu'à des ions $Pb^{2+}$, les autres étant présents sous forme d'atomes d'oxygène pontants (Si-O-Si) et non-pontants (Si-O-Pb). Cela a été confirmé directement par RMN $^{17}O$ dans le cas de verres riches en PbO (60-71 mol%) enrichis en $^{17}O$ (figure 3.13) (Lee et Kim 2015).

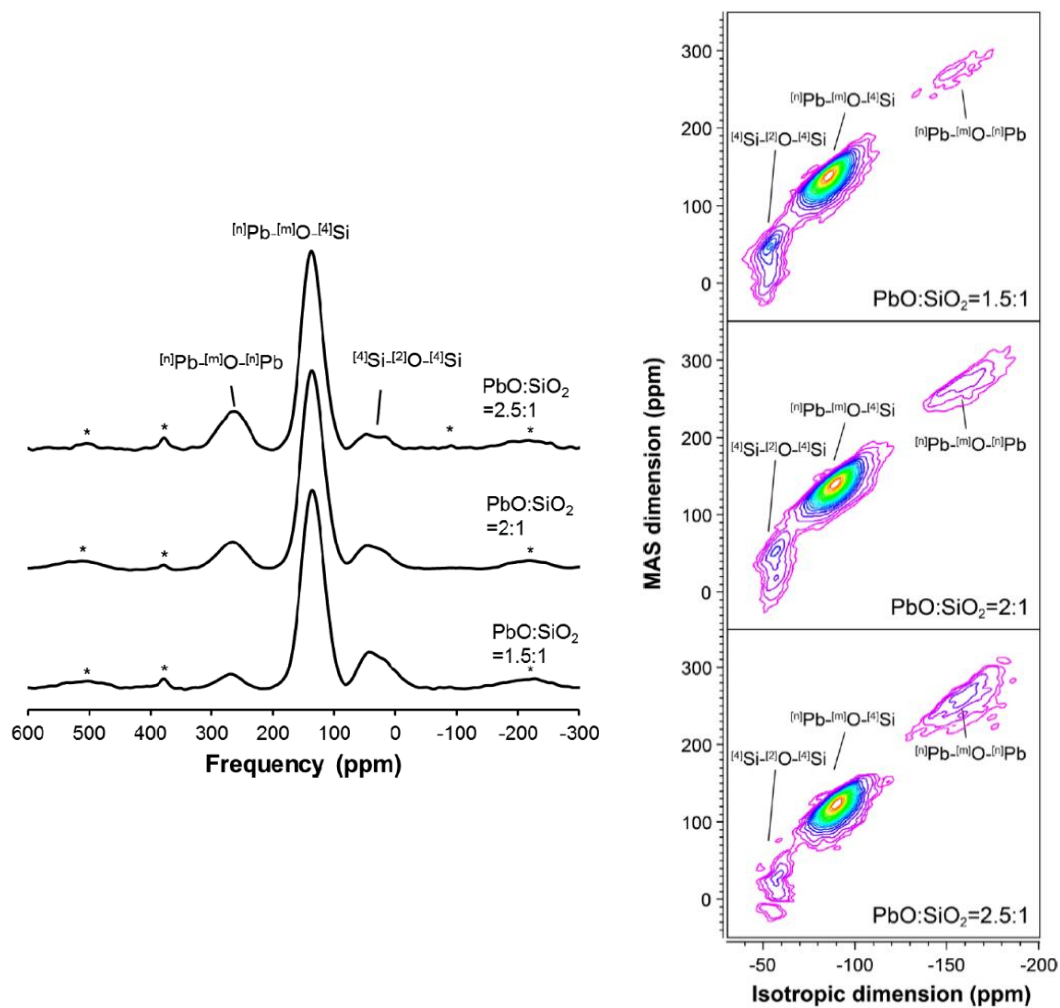

**Figure 3.13.** *(gauche) Spectres RMN MAS $^{17}$O de verres de silicate de plomb enrichis en $^{17}$O. Les pics marqués par des astérisques sont des bandes de rotation. (droite) Spectres RMN 3QMAS des mêmes verres de silicate de plomb enrichis en $^{17}$O. Les ratios des teneurs molaires PbO/SiO$_2$ sont indiqués sur les figures et correspondent à 60 (1.5:1), 66 (2:1) et 71 (2.5:1) mol% PbO. Les différents types d'atomes d'oxygène mis en évidence sont : les atomes d'oxygène pontants ($^{[4]}$Si-$^{[2]}$O-$^{[4]}$Si), les atomes d'oxygène non-pontants ($^{[n]}$Pb-$^{[m]}$O-$^{[4]}$Si) et les atomes d'oxygène connectés uniquement au plomb ($^{[n]}$Pb-$^{[m]}$O-$^{[n]}$Pb). Reproduit avec permission à partir de (Lee S. K., Kim E. J., J. Phys. Chem. C 119, 748-756, 2015). Copyright (2020) American Chemical Society.*

D'autres auteurs ont quant à eux utilisé la spectroscopie Raman pour suivre l'évolution de la structure des verres de silicate de plomb et tenter de quantifier les unités Q$^n$ en réalisant parfois des comparaisons avec les phases cristallines (Furukawa *et al.* 1978 ; Feller *et al.* 2010 ; Kaur *et al.* 2013 ; Ben Kacem *et al.* 2017). Comme en RMN $^{29}$Si, une très forte évolution des spectres Raman avec la teneur en PbO a été observée en accord avec une dépolymérisation croissante du réseau silicaté. De plus, en accord avec les résultats de RMN $^{17}$O (figure 3.13), la formation conjointe de liaisons Pb-O-Pb a été mise en évidence, traduisant une tendance croissante des ions Pb$^{2+}$ à se lier entre eux (clustérisation). Les études réalisées par XPS (O 1s) sur les verres binaires SiO$_2$-PbO ont également mis en évidence une forte diminution de la proportion d'atomes d'oxygène pontants avec la teneur en PbO, cependant la résolution des spectres est insuffisante pour séparer la contribution des atomes d'oxygène non-pontants de celle des atomes d'oxygène non liés au silicium (Smets et Lommen 1982 ; Sorokina *et al.* 1996 ; Dalby *et al.* 2007). Toutefois, certaines de ces études soulignent la persistance d'atomes d'oxygène pontants - c'est-à-dire de liaisons Si-O-Si - dans les compositions les plus riches en PbO (Smets et Lommen1982 ; Dalby *et al.* 2007) en accord avec les études menées par RMN $^{29}$Si (figure 3.12). Signalons pour finir les études réalisées par diffusion des rayons X et des neutrons combinées à de la modélisation structurale par la méthode de Monte Carlo inverse (Suzuya *et al.* 1999 ; Kohara *et al.* 2010) qui ont conduit à une distribution des unités Q$^n$ en fonction de la composition (figure 3.14) cohérente avec les résultats de RMN $^{29}$Si (figure 3.12). En revanche, les modèles structuraux obtenus par dynamique moléculaire et qui ne prennent pas en compte la polarisabilité des ions Pb$^{2+}$ nécessaire pour conduire à des environnements locaux anisotropes donnent des résultats médiocres, ces travaux utilisant des potentiels d'interaction symétriques (Ribycka *et al.* 1999).

L'ensemble de ces résultats relatifs à l'évolution du réseau silicaté des verres SiO$_2$-PbO se rapproche donc qualitativement au moins de l'évolution de la structure des phases cristallines de silicate de plomb présentées dans la section 3.2 (figures 3.7-9). On observe en effet dans les deux cas une forte dépolymérisation du réseau silicaté (vitreux et cristallin) avec la persistance de tétraèdres SiO$_4$ connectés entre eux même pour les teneurs les plus élevées en PbO (figures 3.7, 3.8 et 3.12): unités Q$^1$ et Q$^2$ pour les verres et anneaux Si$_4$O$_{12}$ et unités Q$^1$ respectivement pour les phases cristallines Pb$_2$SiO$_4$ (66 mol% PbO) et Pb$_{11}$Si$_3$O$_{17}$ (78 mol% PbO). Avec également dans les deux cas la présence d'une proportion significative d'atomes d'oxygène uniquement liés au plomb (liaisons Pb-O-Pb) (figures 3.9 et 3.13).

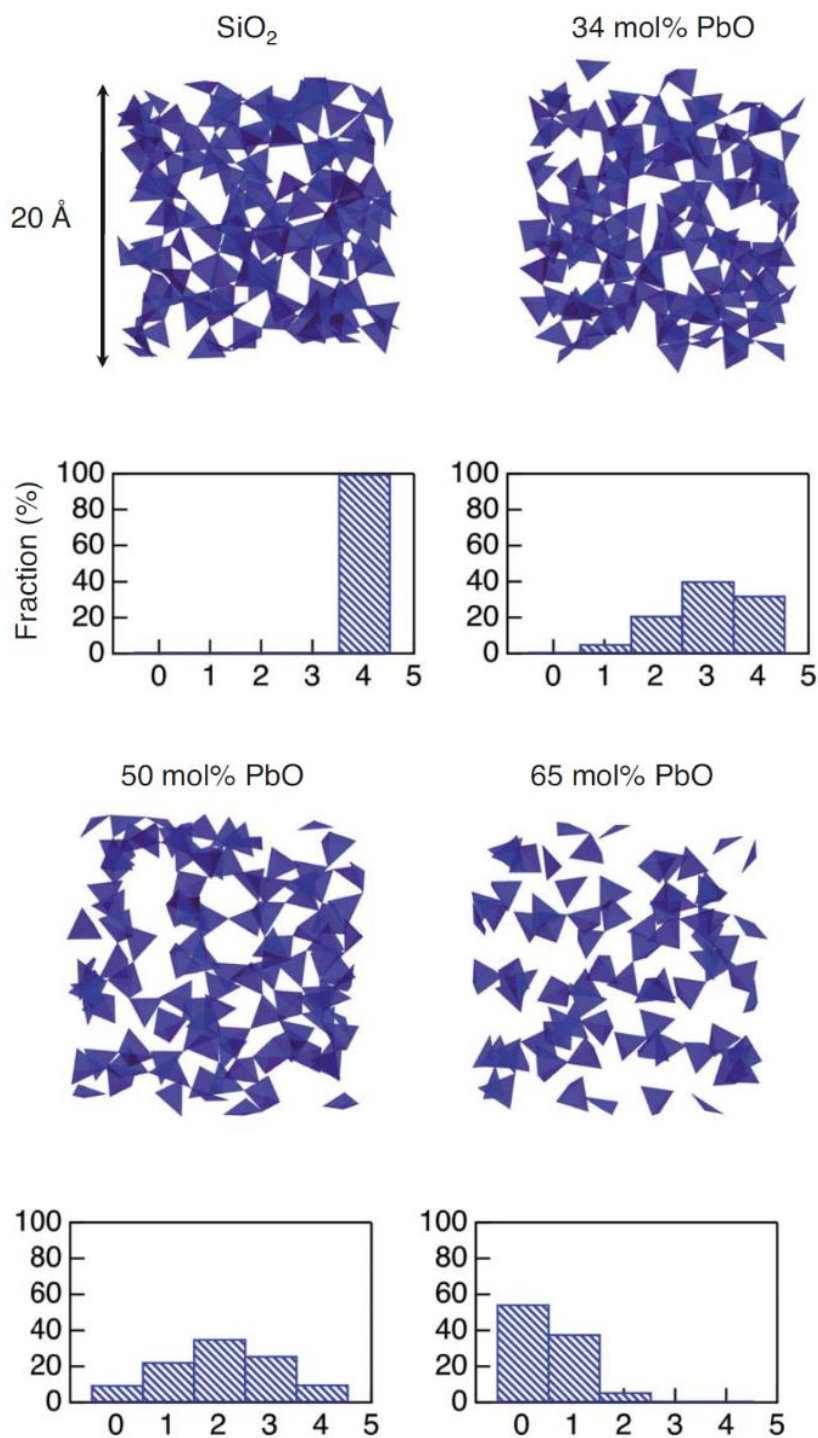

**Figure 3.14.** *Evolution du réseau vitreux silicaté (seuls sont représentés les tétraèdres SiO$_4$) de verres de silicate de plomb et de silice pure en fonction de leur teneur en PbO. Les modèles structuraux ont été obtenus par la méthode de Monte Carlo inverse à partir de résultats de diffusion des rayons X et des neutrons. La distribution des unités Q$^n$ (n = 0,1,2,3,4) déterminée pour les différents verres à partir de ces modèles est également présentée sous forme d'histogrammes (Kohara et al. 2010). Reproduit avec permission à partir de (Kohara, S., Ohno, H., Takata, M., Usuki, T., Morita, H., Suzuya, K., Akola, J., Pusztai, L., Phys. Rev. B, 82, 134209, 2010) Copyright (2020) par l'American Physical Society.*

3.3.3.2.2 Environnement, distribution et rôle structural du plomb

Différentes techniques permettent d'extraire des informations sur l'environnement du plomb dans les verres. Certaines de ces techniques permettent de sonder spécifiquement le plomb, comme les spectroscopies RMN[19], EXAFS et XANES et XPS alors que d'autres techniques comme la diffusion des rayons X et des neutrons bien que donnant des informations précieuses sur l'environnement des ions $Pb^{2+}$ (coordinence, distance Pb-O) sondent également les autres éléments présents dans le réseau vitreux. Les études publiées dans la littérature sur l'environnement et l'incorporation du plomb au moyen de ces techniques d'investigation sont nombreuses et parfois contradictoires, de sorte que le rôle structural du plomb dans les verres silicatés est toujours débattu. Cependant, la majorité des auteurs s'accordent aujourd'hui à dire que les ions $Pb^{2+}$ s'insèrent dans la structure des verres binaires de silicate avec leur paire libre $6s^2$ stéréochimiquement active sous la forme de polyèdres $PbO_n$ très asymétriques et de faible coordinence (n = 3 - 4), de type pyramidal avec des liaisons Pb-O courtes à fort caractère covalent : le plomb aurait donc un environnement local proche de celui qu'il occupe dans les phases cristallines de silicate de plomb et dans l'oxyde de plomb pur (sections 3.2.2 et 3.3.2). De plus, pour les teneurs élevées en PbO, les unités $PbO_n$ tendraient à se lier entre elles pour former des domaines percolants de plus en plus grands jusqu'à constituer l'ossature principale du réseau vitreux, englobant les unités silicatées de plus en plus dépolymérisées ($Q^1$, $Q^0$) comme pour les silicates de plomb cristallisés (figure 3.7). L'existence de tels domaines est en accord avec la mise en évidence par RMN $^{17}O$ d'une proportion croissante d'atomes d'oxygène liés uniquement au plomb (liaisons Pb-O-Pb, figure 3.13). Ces conclusions s'appuient entre autres sur les études d'absorption des rayons X (EXAFS, XANES) et de diffusion des rayons X et des neutrons.

Les études de diffusion des rayons X et des neutrons permettent d'extraire les fonctions de distribution radiales faisant intervenir les différentes paires d'atomes présentes dans la structure et à partir desquelles il est possible de remonter aux distances interatomiques et aux coordinences (figure 3.15). Ainsi Hoppe et al. (Hoppe *et al.* 2003) et Takaishi et al. (Takaishi *et al.* 2005) trouvent des distances Pb-O de l'ordre de 2,30 Å pour les atomes d'oxygène les plus proches avec une coordinence de l'ordre de 3, ces distances et coordinences variant relativement peu avec la teneur en PbO dans les verres (figure 3.15).

---

[19] L'isotope $^{207}Pb$ est le seul isotope stable du plomb (abondance naturelle 22,1%) présentant un spin nucléaire non nul (I = ½), il peut donc être étudié par RMN. Cependant en raison de sa forte polarisabilité, les signaux RMN obtenus sont larges et peu résolus.

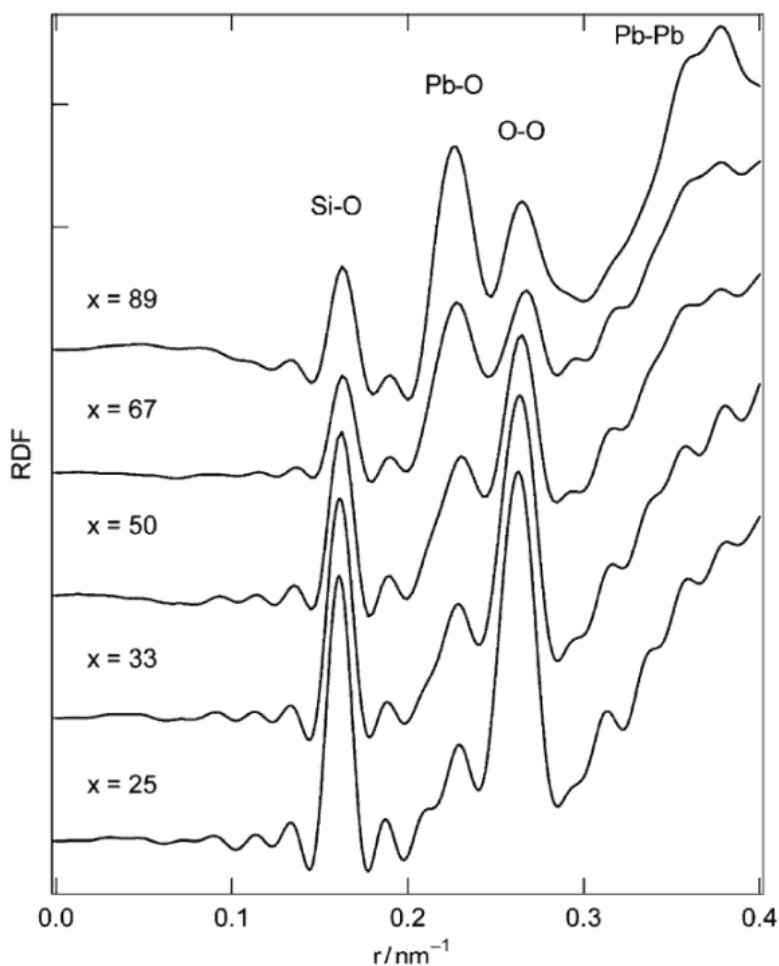

**Figure 3.15.** *Fonctions de distributions radiales calculées à partir des données de diffusion des neutrons pour des verres de composition (1-x)SiO$_2$ - xPbO (x en mol%). Dans le cas de la diffusion des neutrons (montrée ici) on voit bien toutes les corrélations, notamment celles associées aux liaisons Si-O et les O-O car dans ce cas le plomb ne se distingue pas des autres atomes. En revanche, dans le cas de la diffusion des rayons X (non montrée ici) qui est largement dominée par le plomb en raison de son numéro atomique bien plus élevé que celui des autres espèces présentes (Si et O), on obtient essentiellement des informations sur les corrélations Pb-O et Pb-Pb. (Takaishi et al. 2005)*

A partir de leurs résultats de diffusion des rayons X et des neutrons, et en particulier à partir de la distance Pb-Pb qui reste quasi-constante (3,83-3,86 Å) pour l'ensemble des verres étudiés (25-89 mol%), Takaishi et al. ont proposé un modèle structural des verres SiO$_2$-PbO faisant intervenir des unités Pb$_2$O$_4$ constituées de deux pyramides trigonales PbO$_3$ connectées par une arête et orientées dans des directions opposées afin de minimiser la répulsion entre les paires libres 6$s^2$ des deux ions Pb$^{2+}$, ces unités Pb$_2$O$_4$ présentant une charge globale nulle (figure 3.16) (Mizuno *et al.* 2005 ; Takaishi *et al.* 2005). Ces entités seraient présentes dès les plus faibles teneurs en PbO étudiées par les auteurs (25 mol%). Suivant ce modèle, le plomb se comporterait donc en formateur dès les plus faibles concentrations en plomb.

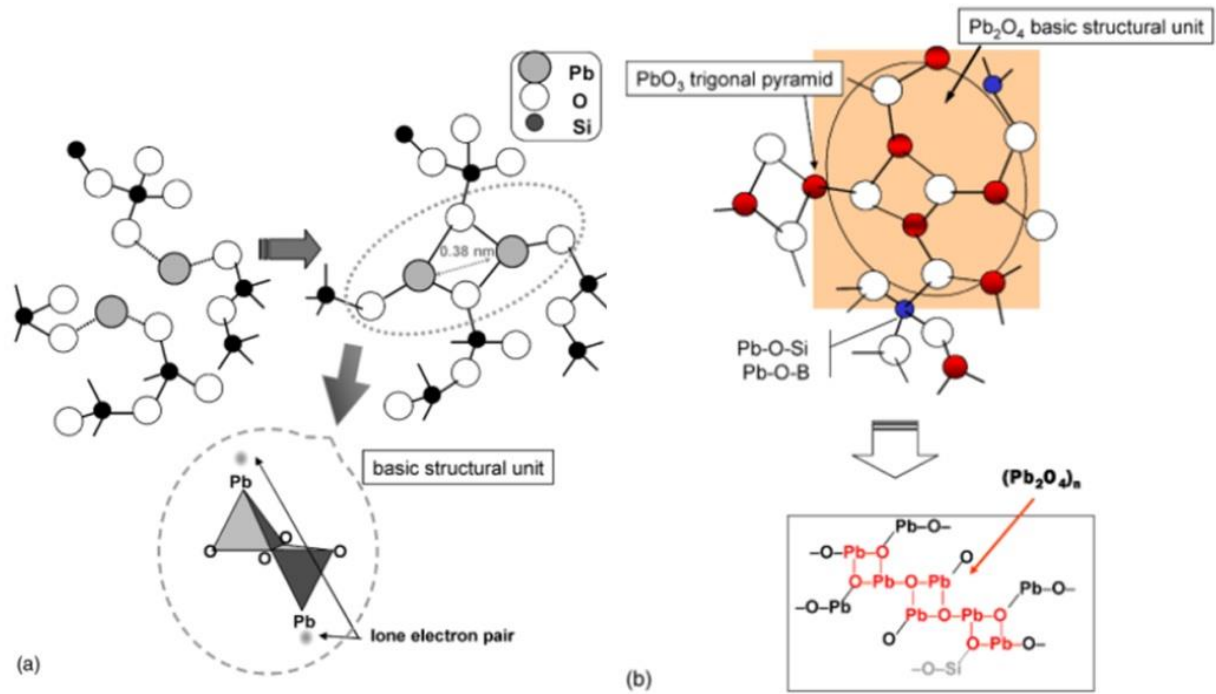

**Figure 3.16.** *Modèle structural proposé par Takaishi et al. (Takaishi et al. 2005) à partir de résultats de diffusion des rayons X et des neutrons. L'unité de base $Pb_2O_4$ constituée de deux pyramides à base triangulaire $PbO_3$ connectées par une arête et avec une disposition opposée des paires libres $6s^2$ du plomb est présentée. (a) pour les faibles teneurs en PbO on voit les unités $Pb_2O_4$ incorporées dans le réseau silicaté, (b) pour les teneurs élevées en PbO on voit la connexion entre les unités $Pb_2O_4$. La distance de 3,8 Å entre les ions $Pb^{2+}$ est indiquée.*

A partir de leurs résultats de diffusion et d'un modèle structural obtenu par simulation avec la méthode de Monte Carlo inverse pour des verres avec des teneurs en PbO comprises entre 30 et 65 mol%, Kohara *et al.* (Kohara *et al.* 2010) n'ont quant à eux pas confirmé l'existence d'unités dimériques $Pb_2O_4$ dans les verres de silicate de plomb, mais un environnement du plomb majoritairement sous forme d'unités $PbO_4$ distordues de type pyramidal à base carrée (figure 3.17), c'est-à-dire un environnement proche de celui du plomb dans la phase cristalline PbO (figures 3.3 et 3.5) et dans les phases cristallines de silicates de plomb (figure 3.7). Au sein de la structure de ces verres, un nombre d'unités $PbO_n$ connectées entre elles croissant fortement avec la teneur en plomb a été mis en évidence. Ces mêmes auteurs ont également présenté une évolution comparée du réseau silicaté et du réseau plombifère en fonction du pourcentage en oxyde de plomb qui montre clairement la prédominance du réseau plombifère à partir d'au moins 50 mol% PbO (figure 3.17). Une étude équivalente menée par Suzuya *et al.* (Suzuya *et al.* 1999) sur la composition à 50 mol% PbO a également conduit ces auteurs à proposer un environnement du plomb sous forme de pyramides PbO$_4$ de forme irrégulière, ces unités formeraient des anneaux constituant le réseau plombifère. Ces modèles faisant intervenir une connectivité croissante d'unités PbO$_4$ au détriment du réseau silicaté se rapproche de celui proposé par plusieurs autres auteurs (Bessada *et al.* 1994 ; Wang et Zhang 1996).

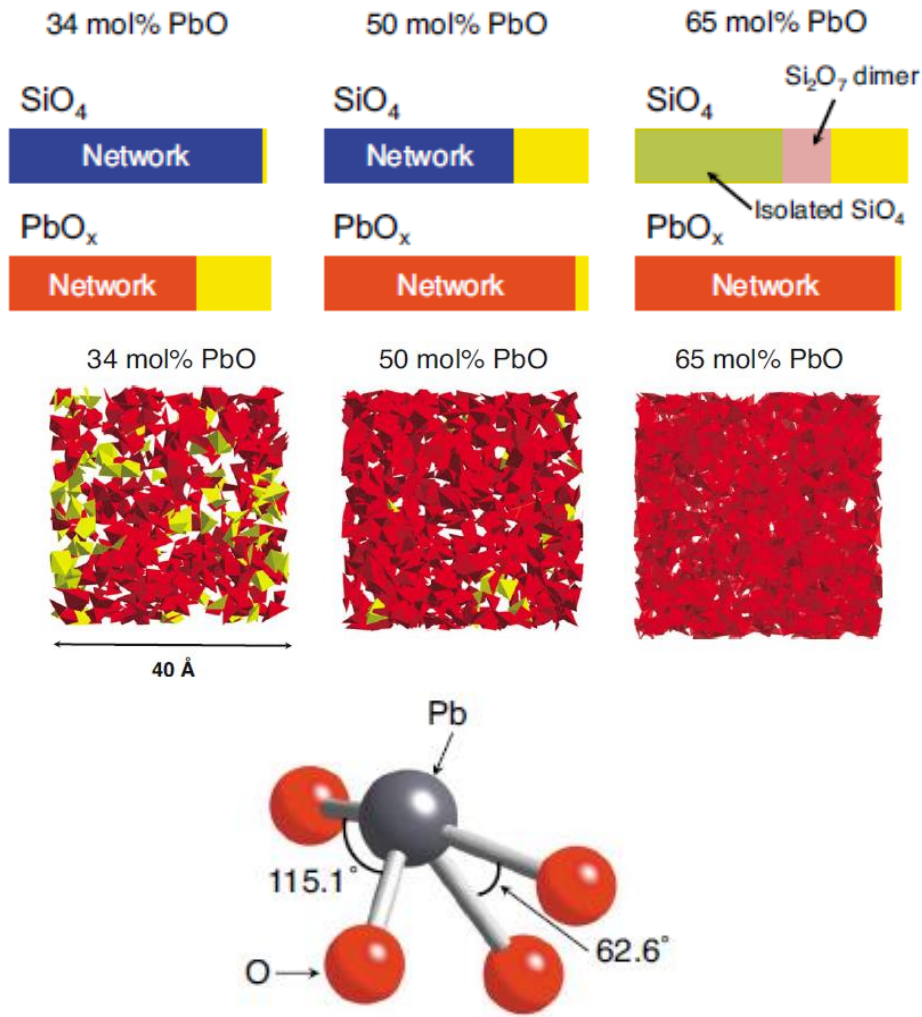

**Figure 3.17.** *Structure locale et étendue de verres de silicate de plomb (30 - 65 mol% PbO) d'après des résultats de diffusion de rayons X et de neutrons et simulation par la méthode de Monte Carlo inverse (RMC). (En bas) Unité pyramidale PbO$_4$ distordue issue des simulations RMC. (En haut) Evolution comparée en fonction de la teneur en plomb des réseaux silicatés (en bleu) et plombifères (en rouge) constitués respectivement d'unités SiO$_4$ et PbO$_x$ connectées entre elles, la proportion d'unités SiO$_4$ et PbO$_x$ ne participant pas à ces réseaux est représentée en jaune, vert et rose. (Au milieu) Evolution du réseau plombifère avec la teneur en plomb (en rouge sont représentées les unités PbO$_x$ connectées entre elles et en jaune les autres, les tétraèdres SiO$_4$ ne sont pas représentés). Reproduit avec*

Plus récemment, Alderman et al. (Alderman 2013 ; Alderman *et al.* 2013) ont proposé une description structurale plus convaincante des verres de silicates de plomb (0 - 80 mol% PbO) également à partir de résultats de diffusion des rayons X et des neutrons mais en utilisant un affinement structural de leurs données expérimentales faisant intervenir des potentiels d'interaction interatomiques empiriques[20] (figures 3.18 et 3.19). Leur modélisation a été optimisée notamment en considérant les ions $Pb^{2+}$ comme des dipôles pour mieux rendre compte de la stéréochimie particulière des sites occupés par ces ions, induite par le doublet $6s^2$ non liant du plomb. Ces auteurs constatent tout d'abord, comme dans les cristaux, une augmentation avec la teneur en PbO de la distance moyenne Si-O imputée à l'évolution de la deuxième sphère de coordination du silicium, et plus particulièrement à l'augmentation du nombre moyen d'ions $Pb^{2+}$ autour des atomes d'oxygène non-pontants. Les auteurs constatent également une légère évolution de l'environnement du plomb entre 35 et 80 mol% PbO : les distances Pb-O courtes ($\leq 2,7$ Å) tendent à devenir plus courtes et plus nombreuses autour des ions $Pb^{2+}$ jusqu'à atteindre quatre liaisons. Inversement, le nombre de distances Pb-O longues (entre 2,7 et 3,27 Å) tend à diminuer. L'environnement des ions $Pb^{2+}$ tendrait donc à évoluer vers celui de la litharge (figure 3.5) lorsqu'on augmente le taux d'oxyde de plomb, et donc lorsqu'on diminue le nombre de silicium dans la seconde sphère de coordination du plomb. Conséquemment, l'environnement de l'oxygène évolue fortement avec le taux d'oxyde de plomb passant d'un environnement $OSi_2$ à un environnement de plus en plus riche en plomb au fur et à mesure de l'addition d'oxyde de plomb, pour tendre vers un environnement $OPb_4$ en moyenne, comme cela a été constaté dans les cristaux. Cela se traduit par la présence, dans tous les verres, de polyèdres $PbO_n$ connectés par arêtes, ce qui permet effectivement d'avoir plusieurs ions $Pb^{2+}$ compensant un atome d'oxygène non-pontant donné et d'avoir plus de deux atomes d'oxygène non-pontants autour d'un ion $Pb^{2+}$, comme c'est le cas dans les cristaux. Cependant, la présence de l'unité $Pb_2O_4$ proposée par Takaishi *et al.* (Takaishi *et al.* 2005) comme motif de base du réseau plombifère indépendamment de la teneur en PbO (figure 3.16) n'a pas été observée dans cette étude. Enfin, une conclusion importante de ces travaux, est la mise en évidence de la formation de zones lacunaires dans les verres riches en oxyde de plomb, dues à l'organisation des doublets non-liants des ions $Pb^{2+}$ (figure 3.19). Cette organisation des doublets non-liants du plomb est rencontrée dans les silicates de plomb cristallisés riches en plomb comme $Pb_{11}Si_3O_{17}$ (figure 3.7), la litharge et le massicot (figure 3.5) qui forment des canaux au sein de la structure cristalline. Ces zones lacunaires ne sont pas présentes dans le verre à faible teneur en oxyde de plomb (figure 3.18) dont la structure est plus proche de celle de l'alamosite $PbSiO_3$ (figure 3.7).

---

[20] Cette technique appelée EPSR (Empirical Potential Structure Refinement) a été mise en œuvre pour étudier l'environnement local autour des ions $Pb^{2+}$, la façon dont les polyèdres $PbO_n$ sont connectés et la disposition des paires libres $6s^2$. Cette méthode qui prend en compte la forte polarisabilité des ions $Pb^{2+}$ est donc plus adaptée que la simple méthode par Monte Carlo inverse mise en œuvre dans les autres études sur les verres de silicate de plomb.

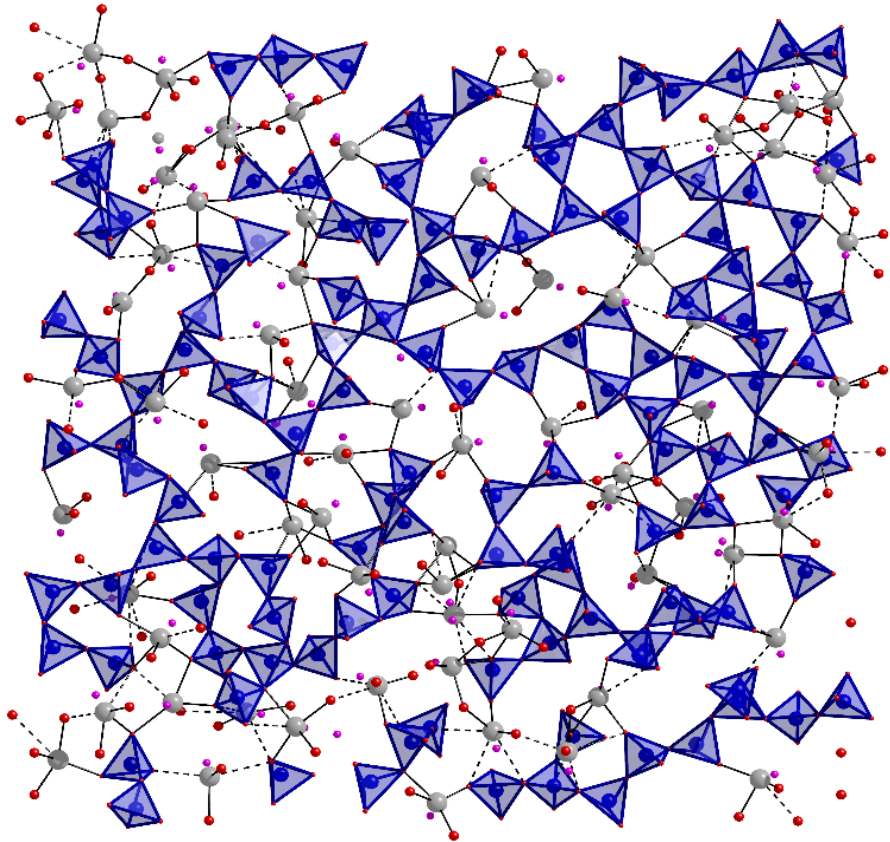

**Figure 3.18.** *Modèle structural proposé par Alderman (Alderman 2013) pour le verre 35PbO·65SiO$_2$ (mol%). Les tétraèdres SiO$_4$ sont représentés en bleu, les ions Pb$^{2+}$par les grosses sphères grises, et les atomes d'oxygène par des petites sphères rouges. Les liaisons Pb-O de longueur inférieure à 2,7Å sont en traits pleins, celles comprises entre 2,7 et 3,27Å sont en pointillés. Les doublets non liants des ions Pb$^{2+}$ sont représentés par de petites sphères roses. Reproduit avec la permission de O. Alderman à partir de (Alderman 2013).*

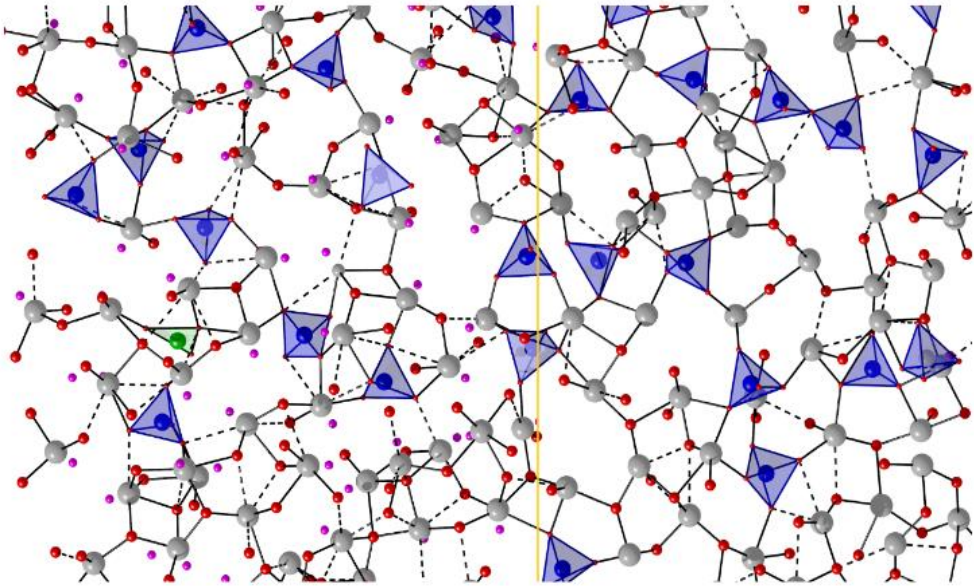

**Figure 3.19.** *Modèle structural proposé par Alderman (Alderman 2013) pour le verre 80PbO·20SiO$_2$. A droite de la séparation verticale (trait jaune), les doublets non liants des ions Pb$^{2+}$ sont omis alors qu'à gauche ils sont représentés (sphères roses). Le triangle vert est un défaut SiO$_3$ provenant du calcul. Reproduit avec la permission de O. Alderman à partir de (Alderman 2013).*

3.3.3.3. *Bilan sur l'évolution structurale des verres binaires SiO$_2$-PbO et rôle du plomb*

L'ensemble des résultats structuraux publiés sur les verres binaires du système SiO$_2$-PbO présentés dans ce chapitre tend donc à mettre en évidence que le rôle structural du plomb ne change pas fondamentalement avec la teneur en PbO et qu'il se rapproche plus d'un rôle d'oxyde formateur. Son environnement à courte distance est très asymétrique et est de type pyramidal déformé à base triangulaire ou carrée (PbO$_3$, PbO$_4$) avec le plomb au sommet et dans tous les cas la paire libre 6$s^2$ des ions Pb$^{2+}$ stréréochimiquement active comme dans les phases cristallines PbO (massicot et litharge) et les silicates de plomb (sections 3.2.2 et 3.3.2). L'environnement du plomb évoluerait donc faiblement en fonction de la teneur en PbO, avec seulement un léger renforcement des liaisons PbO à courte distance, davantage d'environnements tetracoordinés (PbO$_4$) et la présence croissante de connexions Pb-O-Pb avec des liaisons par sommets et arrêtes entre les polyèdres PbO$_n$ formant un réseau plombifère d'importance croissante comme c'est le cas pour les phases cristallines (figures 3.7 et 3.9). De même que pour ces dernières (figures 3.7 et 3.8), une évolution très marquée du réseau silicaté avec la teneur en PbO est observée avec une dépolymérisation progressive (disparition de liaisons Si-O-Si au profit de liaisons Si-O-Pb, figures 3.12 et 3.14), les atomes de silicium ayant un nombre croissant d'ions Pb$^{2+}$ en position de seconds voisins. Cependant, pour des teneurs en PbO supérieures à celle de la composition de l'*oxy-cyclo*silicate de plomb ($Pb_2SiO_4$, 60 mol% PbO) à partir de laquelle on se serait attendu à une dépolymérisation totale du réseau silicaté - comme c'est le cas pour les systèmes de silicates alcalins par exemple - avec seulement des unités $SiO_4$ isolées ($Q^0$), des unités de petites dimensions mais présentant des liaisons Si-O-Si sont toujours présentes même pour les compositions les plus riches en plomb (figures 3.12, 3.13 et 3.14). La présence de ces unités au sein du réseau plombifère pour les compositions riches en plomb, ainsi que l'existence d'une distribution d'environnement pour les ions $Pb^{2+}$ et de connexions Pb-O-Pb et Pb-O-Si au sein de la structure des verres (et donc des liquides correspondants au moment de la fusion et avant trempe) doit donc contribuer à freiner la cristallisation du liquide surfondu au moment de la trempe et doit donc faciliter l'obtention d'un verre. En revanche dans le cas des cations alcalins et alcalino-terreux qui sont liés aux anions $O^{2-}$ par des liaisons uniquement ioniques et non-directionnelles et qui présentent des environnements assez sphériques dans les verres de silicate, on comprend bien que la réorganisation du liquide pour conduire à des cristaux est facilitée et qu'il n'est donc pas possible de vitrifier des liquides silicatés pour des teneurs élevées en oxydes alcalins et alcalino-terreux. L'ensemble de ces résultats structuraux permet en grande partie d'expliquer la variation des propriétés physiques des verres binaires $SiO_2$-PbO en fonction de la concentration en PbO dont nous avons parlé dans la section 3.3.3.1. Ainsi la dépolymérisation du réseau silicaté, le fait que la force moyenne des liaisons entre l'oxygène et ces voisins Si et Pb diminue avec la teneur en plomb (les liaisons Pb-O étant plus faibles que les liaisons Si-O) et la forte polarisabilité des ions $Pb^{2+}$ (et dont leur forte déformabilité) permettent de comprendre la diminution avec la teneur en PbO de la viscosité du liquide, de la température de transition vitreuse et de la résistivité électrique (figure 3.10). De même l'augmentation du coefficient de dilatation des verres $SiO_2$-PbO avec la teneur en PbO peut s'expliquer par une diminution de la force moyenne des liaisons comme pour les verres de silicate alcalins, cependant les liaisons Pb-O ayant un caractère iono-covalent à la différence des liaisons entre les ions alcalins $R^+$ et les anions $O^{2-}$ qui sont purement ioniques, le coefficient de dilatation des verres de silicate alcalins demeure toujours supérieur à celui des verres de silicate de plomb pour une même teneur molaire en $R_2O$ et PbO (Scholze 1980). Pour finir, la diminution du volume molaire des verres $SiO_2$-PbO lorsque la concentration en plomb augmente pour les plus faibles teneurs en PbO (Shelby 1983 ; Kaur *et al.* 2013 ; Ben Kacem *et al.* 2017) pourrait s'expliquer par une diminution du volume libre existant dans les verres de silicate riches en $SiO_2$ suite à la coupure progressive des liaisons Si-O-Si. Alors que la stabilisation du volume libre voire sa réaugmentation à partir de 50 mol% PbO pourrait être expliquée par la formation d'un réseau plombifère de plus en plus étendu et par la présence d'un nombre croissant de paires $6s^2$ libres associées au plomb créant des zones lacunaires dans la structure (figure 3.19). Il est important de signaler que la forte polarisabilité de l'ion $Pb^{2+}$ et l'existence de la paire libre $6s^2$ stéréochimiquement active pourrait être à l'origine de la bien meilleure solubilité d'oxydes métalliques comme $Cr_2O_3$ dans les verres silicatés renfermant de l'oxyde de plomb par rapport aux verres silicatés classiques renfermant des oxydes alcalins et alcalino-terreux et dans lesquels l'oxyde de chrome est peu soluble (Ashour et Shehata 1972 ; Volf 1984).

## 3.4. Les verres du système $SiO_2$-PbO-$R_2O$ (R= Na, K)

### 3.4.1. *Généralités sur les verres ternaires $SiO_2$-PbO-$R_2O$*

Après les verres binaires $SiO_2$-PbO, intéressons-nous à la structure et aux propriétés des verres ternaires $SiO_2$-PbO-$R_2O$ où R représente un alcalin tels que Na ou K. Ces verres ternaires présentent de nombreuses applications dans les domaines par exemple de la verrerie de luxe (cristallerie), des glaçures et émaux et des verres absorbant les rayonnements ionisants comme on l'a vu dans le chapitre 1. Ainsi le verre de cristal appartient au ternaire $SiO_2$-PbO-$K_2O$ avec des teneurs molaires par exemple de l'ordre de 77% $SiO_2$, 11% $K_2O$ et 10% PbO (avec suivant les productions, un peu de $Na_2O$, de $Al_2O_3$ et de traces d'agents d'affinage tel que $Sb_2O_3$), ce qui conduit à des teneurs massiques en PbO de l'ordre de 28% dans ce cas. Alors que par exemple dans le cas des glaçures du patrimoine mixtes alcalines plombifères opacifiées à l'étain on trouve des teneurs massiques de l'ordre de 20-40% PbO et 5-12% $Na_2O$ + $K_2O$ à comparer aux teneurs de l'ordre de 45-60% PbO et moins de 2% $Na_2O$ + $K_2O$ pour les glaçures transparentes plombifères (Tite *et al.* 1998, chapitre 10). Dans le cas du verre de cristal, l'intérêt de l'ajout de plomb au verre binaire de silicate de potassium est multiple (fusion, mise en forme, usinage, propriétés optiques, nappage des glaçures, voir chapitre 1). Parallèlement, l'ajout de $K_2O$ au verre binaire $SiO_2$-PbO permet de réduire la tendance à la démixtion des verres de silicate de plomb présentant des teneurs en PbO faibles ou modérées (< 35 mol%) (section 3.3.3.1) (Calvert et Shaw 1970 ; Vogel 1994) et facilite également la fusion du mélange vitrifiable. Etant donné l'importance tant au niveau industriel que patrimonial des verres ternaires $SiO_2$-PbO-$R_2O$, de nombreuses études portant sur leur résistance hydrolytique en milieu immergé ou en atmosphère humide ont été réalisées, dans un but à la fois sanitaire (Angeli *et al.* 2016) et de conservation (Palomar *et al.* 2020). L'altération de ce type de verres est largement abordée dans le chapitre 13.

### 3.4.2. *Structure des verres ternaires $SiO_2$-PbO-$R_2O$*

La structure des verres binaires $SiO_2$-PbO en fonction de leur teneur en PbO a été largement présentée dans la section 3.3.2. Nous allons maintenant considérer l'effet sur la structure du verre et l'environnement du plomb de la présence simultanée de PbO et d'un oxyde alcalin (tel que $Na_2O$ ou $K_2O$). Les oxydes alcalins sont bien connus pour leur rôle de modificateur dans les verres silicatés binaires $SiO_2$-$R_2O$, conduisant à une dépolymérisation plus ou moins marquée du réseau silicaté en fonction de la teneur en $R_2O$ (Maekawa *et al.*1991). Il est intéressant de se pencher ici sur la façon dont la structure des verres binaires de silicate de plomb est affectée quand un oxyde modificateur (ici $R_2O$) est ajouté à leur composition et en particulier sur la façon dont l'environnement et la distribution des ions $Pb^{2+}$ peuvent être modifiés. Cependant, à la différence des études structurales réalisées sur les verres binaires de silicate de plomb pour lesquels la littérature est particulièrement abondante, celle relative aux verres ternaires $SiO_2$-PbO-$R_2O$ est moins riche malgré les applications industrielles importantes de ces verres. Plusieurs méthodes spectroscopiques ont été mises en œuvre pour étudier la structure de ces verres telles que les spectroscopies Raman, IR, XPS.

Parallèlement, alors que les phases cristallines du système PbO-$SiO_2$ ont fait l'objet d'une étude structurale quasi-systématique (section 3.3.2), on ne sait que très peu de choses des composés cristallins plus complexes tels que ceux des ternaires $SiO_2$-PbO-$R_2O$. Ainsi, en dépit du nombre considérable de compositions possibles, les silicates de plomb cristallisés contenant des cations alcalins ou alcalino-terreux n'ont fait l'objet d'études structurales que dans le cadre de travaux aussi rares qu'anciens portant sur le disilicate $K_2Pb_2Si_2O_7$ (McMurdie 1941) et son isotype $K_2Pb_2Ge_2O_7$ (Bassi *et al.* 1965), le cyclosilicate $PbCa_2Si_3O_9$ ou *margarosanite* (Freed *et al.* 1969) et l'oxy-monosilicate $Li_{10}PbO_2(SiO_4)_2$ (Brandes *et al.* 1994). Malgré des teneurs en plomb disparates, ces structures montrent que l'effet dépolymérisant des oxydes alcalins et alcalino-terreux - à des degrés divers en fonction de leur teneur - contribue à favoriser comme pour les silicates de plomb binaires des environnements asymétriques pour les ions $Pb^{2+}$ (paire libre $6s^2$ stéréochimiquement active) et des pontages Pb-O-Pb forts. Cet effet est particulièrement marqué pour la phase cristalline $Li_{10}PbO_2(SiO_4)_2$ qui présente des chaines infinies -Pb-O- grâce à son taux exceptionnellement élevé d'alcalins qui dépolymérise totalement le réseau silicaté.

#### 3.4.2.1. *Le réseau vitreux silicaté*

Plusieurs études sur les verres ternaires $SiO_2$-PbO-$R_2O$ ont été reportées dans la littérature en faisant varier les proportions relatives des trois oxydes. Ainsi, en ajoutant des teneurs croissantes en $Na_2O$ à un verre de composition 50 $SiO_2$ - 50 PbO (mol%) ou des teneurs croissantes en PbO à un verre de composition 75 SiO$_2$ - 25 Na$_2$O (mol%), des études réalisées par RMN $^{29}$Si (Shrikhande *et al.* 2001 ; Shrikhande *et al.* 2007) ont clairement mis en évidence une dépolymérisation du réseau silicaté c'est-à-dire une diminution de la proportion d'atomes d'oxygène pontants connectant les atomes de silicium, comme c'est le cas lors de l'ajout de teneurs croissantes en PbO à SiO$_2$ (figure 3.12). Une évolution structurale similaire a été observée lors de la substitution progressive de SiO$_2$ par PbO (jusqu'à 50 mol%) dans un verre de cristal industriel de composition proche de 77,9 SiO$_2$ - 10,7 PbO - 11,4 K$_2$O (mol%) comme cela est présenté dans la figure 1 du chapitre 13. Pour ce dernier verre, la RMN $^{29}$Si a par ailleurs montré qu'il était essentiellement constitué d'unités Q$^3$ (59%) et Q$^4$ (41%), ce qui correspond à 25,7 % d'atomes d'oxygène non-pontants et est en parfait accord à ce qu'on obtient par calcul à partir de sa composition en supposant que K$_2$O et PbO transfèrent totalement leurs anions O$^{2-}$ au réseau silicaté (Angeli *et al.* 2016) (réaction acido-basique totale). Cela pourrait laisser penser que ces deux oxydes ont des effets similaires sur le réseau silicaté et un rôle structural similaire. Cependant, pour des teneurs en PbO plus élevées (> 30 mol%), un désaccord croissant est mis en évidence entre le résultat du calcul de la proportion d'atomes d'oxygène non-pontants réalisé à partir de la composition du verre et les résultats RMN (chapitre 13). Cela suggère fortement que pour des teneurs en plomb assez élevées, une partie de l'oxygène apportée par PbO se retrouve sous forme de liaisons Pb-O-Pb, comme c'est le cas pour les verres binaires SiO$_2$-PbO riches en plomb (section 3.3.3.2). Une étude réalisée par XPS (O 1s) sur des verres ternaires en substituant progressivement PbO par Na$_2$O va également dans ce sens (Gee *et al.* 2001).

3.4.2.2. *Environnement et distribution du plomb et des alcalins dans le réseau vitreux*

A la différence des verres binaires SiO$_2$-PbO, aucune donnée d'EXAFS, de XANES, de diffusion des rayons X ou des neutrons ni de RMN $^{207}$Pb n'est à notre connaissance disponible dans la littérature pour les verres ternaires SiO$_2$-PbO-R$_2$O qui aurait pu apporter des informations directes sur l'environnement du plomb au sein de ces verres. La situation est donc différente des verres binaires SiO$_2$-PbO pour lesquels il a été mis en évidence par ces techniques que le plomb sous forme d'ions Pb$^{2+}$ (avec sa paire 6*s*$^2$ libre stéréochimiquement active) était localisé dans des sites très asymétriques connectés au réseau silicaté par des liaisons Si-O-Pb assez fortes, ces sites pouvant également être connectés entre eux pour les teneurs élevées en PbO. Cependant, par analogie avec les phases cristallines ternaires de silicates de plomb alcalins ou alcalino-terreux évoquées plus haut, il est fortement probable que dans les verres ternaires des mêmes systèmes les ions Pb$^{2+}$ soient également localisés dans des environnements asymétriques avec leur paire libre 6*s*$^2$ stéréochimiquement active. De plus, une étude récente menée par RMN sur des verres ternaires SiO$_2$-PbO-K$_2$O enrichis en $^{17}$O a apporté de riches informations sur la façon dont les ions $Pb^{2+}$ se distribuent par rapport aux ions $K^+$ dans la structure du réseau vitreux en fonction de la teneur en PbO (substitué à $SiO_2$) (figure 3.20) (Angeli *et al.* 2016). Cette étude se rapproche de celle de Lee *et al.* (Lee *et al.* 2015) menée sur les verres binaires $SiO_2$-PbO (figure 3.13) et est détaillée dans le chapitre 13 (figure 3.2). Elle montre comment évolue la proportion des différents types de connections entre Si, Pb, O et K dans la structure du verre en fonction de sa teneur en plomb : Si-O-Si, Si-O-K, Si-O-Pb, Si-O-(Pb,K) et Pb-O-Pb. Il apparaît que pour les teneurs les plus faibles en PbO telle que celle du verre de cristal (~ 11 mol%), les ions $Pb^{2+}$ partagent avec une partie des ions $K^+$ les atomes d'oxygène non-pontants de la structure sous la forme de liaisons mixtes Si-O-(Pb,K), une part significative des ions $K^+$ restant cependant connectés seuls à des atomes d'oxygène non-pontants sous la forme de liaisons Si-O-K (figure 3.20).

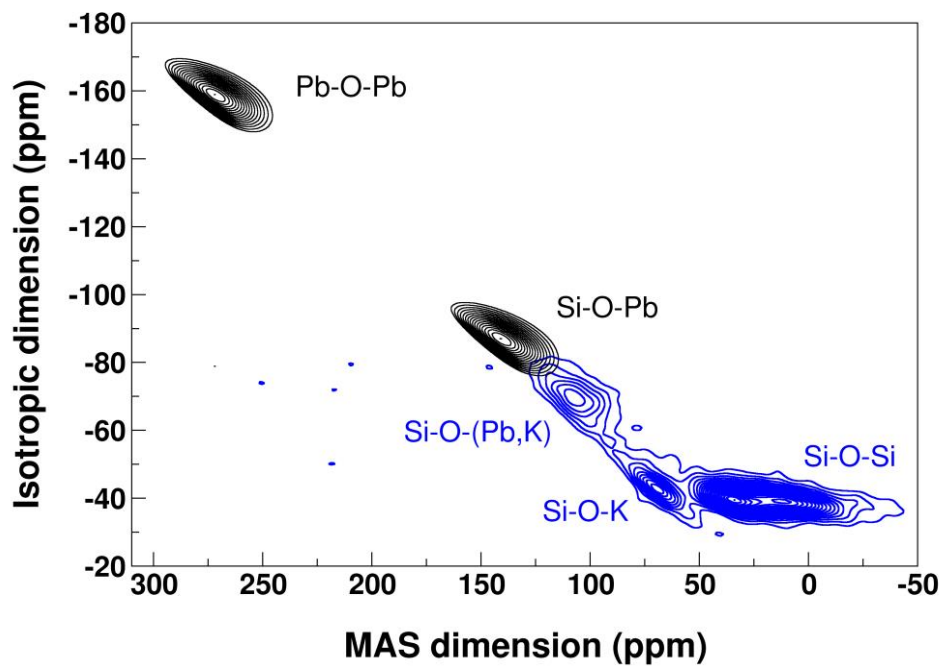

**Figure 3.20.** *Spectre $^{17}O$ MQ-MAS d'un verre de cristal (77,1 $SiO_2$ - 10,6 PbO - 11,3 $K_2O$ - 0,8 $Na_2O$ - 0,2 $Sb_2O_3$ mol%) en bleu montrant la contribution des atomes d'oxygène participant aux liaisons Si-O-Si, Si-O-K et Si-O-(Pb,K). Pour comparaison la position de la contribution des liaisons Si-O-Pb et Pb-O-Pb obtenue par Lee et al. (Lee et al. 2014) pour des verres binaires $SiO_2$-PbO riches en PbO est reportée en noir sur le spectre. Reproduit avec permission à partir de (Angeli F., Jollivet P., Charpentier T., Fournier M., Gin S., Environ. Sci. Technol. 50, 11549-11558, 2016). Copyright (2020) American Chemical Society.*

Il est important de noter que pour ces teneurs en PbO la présence d'atomes d'oxygène isolés du réseau silicaté sous forme de liaisons Pb-O-Pb n'est pas observée, ce qui montre que l'oxyde

PbO réagit totalement avec SiO$_2$ en accord avec les résultats de RMN $^{29}$Si évoqués plus haut. En revanche lorsque la teneur en PbO augmente jusqu'à au moins 30 mol%, la totalité des ions K$^+$ se retrouvent à partager les atomes d'oxygène non-pontants avec les ions Pb$^{2+}$. C'est seulement pour des teneurs plus élevées en PbO (> 30 mol%) que sont mises en évidence des connections Pb-O-Pb comme le suggéraient les études de RMN $^{29}$Si et XPS (O 1s) évoquées dans la section 3.4.2.1. Ce qui va dans le sens de la clustérisation croissante des entités PbO$_n$ et de la formation progressive d'un réseau plombifère comme pour les verres binaires de silicate de plomb et est en accord avec la structure des phases cristallines ternaires de silicates alcalins ou alcalino-terreux riches en plomb évoquées plus haut. Cela a été confirmé par une étude par dynamique moléculaire et ajustement des résultats RMN par la méthode de Monte Carlo inverse (chapitre 13, figure 4). En outre, des connections Si-O-Pb sans ions K$^+$ dans leur voisinage immédiat doivent être de plus en plus abondantes. Le changement important de l'environnement des ions potassium avec la teneur en plomb mise en évidence pour ces verres est en accord avec la nette évolution du spectre XPS (Na 1s) observée avec la teneur en plomb pour les verres 70 SiO$_2$ - x PbO - (30-x) Na$_2$O (mol%) (Gee *et al.* 2001). Ces évolutions structurales, et en particulier la formation de liaisons mixtes Si-O-(Pb,K) aux dépens de liaisons simples Si-O-K, permettent de comprendre l'effet de la substitution partielle de SiO$_2$ par PbO sur la résistivité électrique (augmentation) des verres de silicate alcalins (Scholze 1980), les gros ions Pb$^{2+}$ gênant la mobilité des ions alcalins.

## 3.5. Les verres du système SiO$_2$-PbO-Al$_2$O$_3$

### 3.5.1. *Généralités sur les verres ternaires SiO$_2$-PbO-Al$_2$O$_3$*

Bien que les verres du système ternaire SiO$_2$-PbO-Al$_2$O$_3$ trouvent des applications anciennes par exemple comme glaçures déposées à la surface des objets en céramique[21] (Tite *et al.* 1998) et beaucoup plus récemment comme matériaux renfermant des ions terres rares pour l'optique et la dosimétrie (Bhargavi *et al.* 2012 ; Rao *et al.* 2013 ; Bhargavi *et al.* 2014 ; Rao *et al.*2 014 ; Bhargavi *et al.* 2015), ainsi que pour la protection contre les rayonnements ionisants (Singh *et al.* 2014), seul un nombre limité de travaux ont été publiés sur les verres de ce système. C'est le cas en particulier des études réalisées sur leur structure et sur l'impact de la présence d'alumine sur l'environnement des ions Pb$^{2+}$ au sein du réseau vitreux, en comparaison de l'ampleur des études réalisées sur les verres binaires SiO$_2$-PbO (section 3.3.3). Parmi celles-ci on peut citer essentiellement les études récentes réalisées par Bhargavi *et al.* (Bhargavi *et al.* 2012 ; Bhargavi *et al.* 2014 ; Bhargavi *et al.* 2015), Rao *et al.* (Rao *et al.* 2013 ; Rao *et al.* 2014) et Ben Kacem (Ben Kacem 2017) au moyen des spectroscopies vibrationnelles ou optiques, et par Roisine (Roisine 2018) à l'aide des spectroscopies RMN et Raman et de la diffusion des rayons X et des neutrons, et celles plus anciennes de Prabakar et Rao (Prabakar et Rao 1991) menées par spectroscopie RMN et de Roy (Roy 1990) conduite par spectroscopie infra-rouge. Notons toutefois que d'autres études structurales ont été réalisées sur des verres plus complexes du système $SiO_2$-$PbO$-$Al_2O_3$-$B_2O_3$ renfermant du bore en plus de l'aluminium, mais elles ne seront pas abordées ici (Sawvel *et al.* 2005 ; Saini *et al.* 2009 ; Falconeri *et al.* 2012). Bien que l'extension du domaine vitrifiable au sein du ternaire $SiO_2$-$PbO$-$Al_2O_3$ n'ait pas été reportée dans la littérature, on peut cependant relever que la teneur en $Al_2O_3$ des verres étudiés n'excède jamais 30 mol% (Ben Kacem 2017) et que des verres avec seulement 10 mol% de $SiO_2$ (10 $SiO_2$ - 80 $PbO$ - 10 $Al_2O_3$) ont pu être obtenus par fusion-trempe classique (Prabakar et Rao 1991). Il est intéressant de préciser qu'il est même possible de préparer par hypertrempe des verres binaires $Al_2O_3$-$PbO$ sans silice avec des teneurs en $PbO$ comprises entre 50 et 95 mol% (Kantor *et al.* 1973 ; Morikawa *et al.* 1981).

---

[21]Dans le cas des glaçures plombifères très riches en plomb, l'alumine présente dans leur composition (< 10 % en masse Al$_2$O$_3$) (Tite *et al.* 1998 ; Roisine 2018) peut provenir soit des matières premières déposées à la surface des objets en céramique avant cuisson, soit de la dissolution partielle dans la fine couche de liquide en fusion riche en SiO$_2$ et en PbO d'une fraction de l'alumine présente initialement dans la pâte constituant les objets.

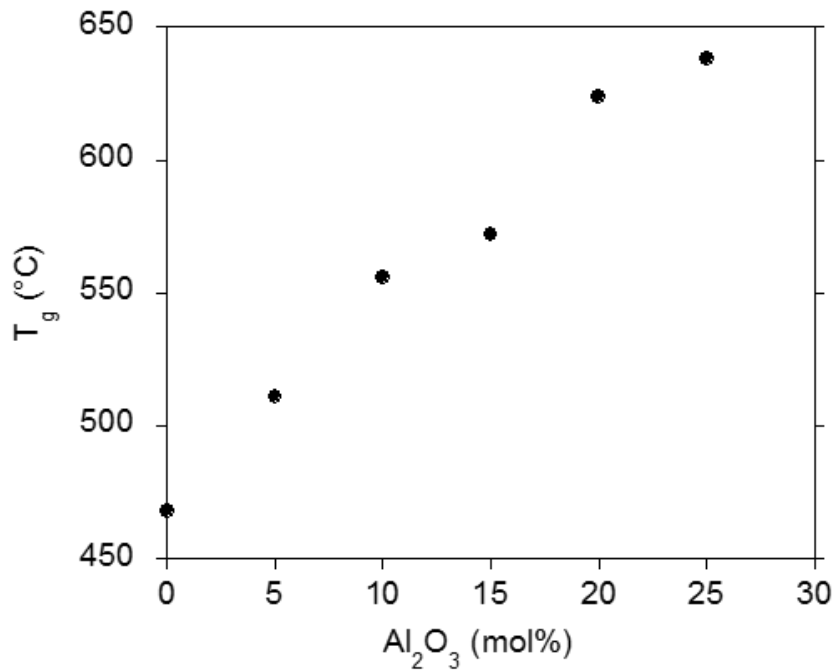

**Figure 3.21.** *Evolution de la température de transition vitreuse $T_g$ en ajoutant des teneurs croissantes en $Al_2O_3$ à la composition 40$SiO_2$ - 60$PbO$ (mol%) (d'après les données de Roisine (Roisine 2017)).*

L'intérêt d'ajouter de l'aluminium à la composition des verres de silicate de plomb est multiple en raison de l'effet important de cet élément sur leurs propriétés. Ainsi, $Al_2O_3$ permettrait d'accroître la résistance mécanique des verres binaires du système $SiO_2$-PbO et de diminuer leur tendance à la cristallisation (Mylyanych *et al.* 1999). De plus, d'après le diagramme de phase du système $SiO_2$-PbO-$Al_2O_3$ (Chen *et al.* 2001), l'introduction d'alumine dans le binaire $SiO_2$-PbO réduirait la tendance à la démixtion du liquide. Cependant, la température de liquidus tend à fortement augmenter, notamment pour les compositions les plus riches en plomb correspondant en particulier au domaine des glaçures plombifères, ce qui peut poser des problèmes de fusion et limite donc la quantité d'alumine qu'on peut y ajouter. De plus, pour ce type de compositions riches en PbO, la température de transition vitreuse et la viscosité du liquide augmentent rapidement avec le taux d'alumine (figures 3.21et 3.22).

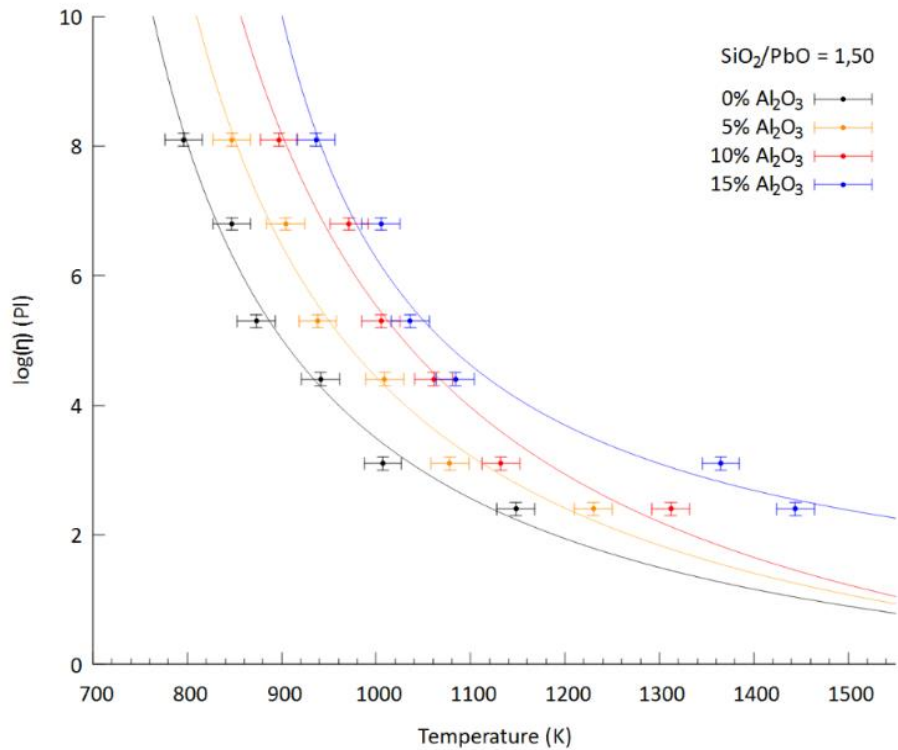

**Figure 3.22.** *Evolution de la viscosité $\eta$ (en poiseuilles Pl) en fonction de la température suite à l'ajout de teneurs croissantes en $Al_2O_3$ à la composition $40SiO_2$ - $60PbO$ (mol%)(Roisine 2017).*

Remarquons également l'existence d'un grand nombre de phases cristallines ternaires (11 phases selon Chen *et al.* (Chen *et al.* 2001)) dans ce domaine de composition qui pourraient contribuer, en plus de la forte augmentation de la viscosité du liquide en fusion, à faciliter sa vitrification au cours du refroidissement selon le principe de « frustration » (Mysen et Richet 2005). Signalons pour finir l'effet bénéfique de la présence d'aluminium dans les verres de silicate de plomb renfermant des ions terres rares conduisant à une meilleure dispersion de ces derniers dans la structure vitreuse (destruction des clusters) (Rao *et al.* 2013).

### 3.5.2. Structure des verres ternaires $SiO_2$-$Al_2O_3$-$PbO$

#### 3.5.2.1. *Les phases cristallines du système $SiO_2$-$Al_2O_3$-$PbO$*

De même que pour les systèmes ternaires de silicates de plomb alcalins ou alcalino-terreux évoqués plus haut (section 3.4.2), les structures connues des oxydes ternaires Pb-Si-Al-O sont très peu nombreuses et se résument au feldspath de plomb $PbAl_2Si_2O_8$ (Benna *et al.* 1996) et à des édifices ouverts de type zéolite (Ronay et Seff 1985 ; Yeom *et al.* 1997). Cependant, à la différence des phases cristallines des systèmes $SiO_2$-$PbO$, $SiO_2$-$PbO$-$R_2O$ et $SiO_2$-$PbO$-$R'O$ (R : alcalin et R' : alcalino-terreux), dans ces structures les longueurs de liaisons Pb-O sont relativement régulières. Leur environnement dans $PbAl_2Si_2O_8$ est ainsi très semblable à celui des ions $Sr^{2+}$ dans le feldspath homologue mettant ainsi en évidence l'absence d'activité stéréochimique de la paire $6s^2$ du plomb dans cette phase très riche en aluminium. Cela s'explique vraisemblablement par la force du squelette aluminosilicate et sa réticulation maximale. Pontant deux tétraèdres $SiO_4$ et $AlO_4$, un anion oxyde ne dispose plus en effet que de 0,25 u.v. environ à partager avec les ions $Pb^{2+}$ voisins, ce qui interdit la formation de liaisons fortes, et donc les pontages Pb-O-Pb fortement structurants observés dans les silicates dépolymérisés des systèmes $SiO_2$-$PbO$, $SiO_2$-$PbO$-$R_2O$ et $SiO_2$-$PbO$-$R'O$. Au sein de ce type de structure cristalline riche en aluminium, les ions $Pb^{2+}$ jouent un rôle de compensateur de charge des unités tétraédriques $AlO_4^-$, à la différence des systèmes précédents considérés dans ce chapitre.

#### 3.5.2.2. *Le réseau vitreux aluminosilicaté et l'environnement du plomb*

Dès les études structurales les plus anciennes menées sur les verres ternaires $SiO_2$-$Al_2O_3$-$PbO$ (Prabakar et Rao 1991), il est apparu par RMN $^{27}Al$ que l'aluminium s'incorporait dans leur structure essentiellement sous la forme d'entités tétraédriques $AlO_4^-$ et qu'une partie des ions $Pb^{2+}$ présents dans la structure agissaient donc en tant que compensateurs de charge de ces entités, cela pour des teneurs en $Al_2O_3$ de 10 mol% et en PbO comprises entre 30 et 80 mol%. Concernant le réseau silicaté, les études menées plus récemment sur différentes séries de verres ternaires par spectroscopies Raman et RMN $^{29}Si$ ont montré que l'ajout d'alumine ou sa substitution partielle à PbO dans les verres binaires $SiO_2$-$PbO$ conduisaient dans tous les cas à une forte évolution de leur structure (Ben Kacem 2017 ; Roisine 2018) (figures 3.23 et 3.24).

Ainsi, en ajoutant des quantités croissantes d'alumine aux verres de silicate de plomb, on observe un décalage vers les basses énergies et un changement de forme de la bande Raman comprise entre 800 et 1200 cm$^{-1}$ associée aux vibrations d'élongation des liaisons T-O des entités tétraédriques $TO_4$ (T = Si, Al) (figure 3.23). L'évolution de cette bande ne peut s'expliquer que par l'incorporation progressive d'aluminium sous forme d'unités $AlO_4^-$ dans le réseau silicaté et non par une dépolymérisation de ce dernier d'après les résultats de RMN $^{17}$O (figures 3-25 et 3-26) et l'évolution des propriétés physiques (forte augmentation de $T_g$ et de la viscosité, figures 3.21 et 3.22).

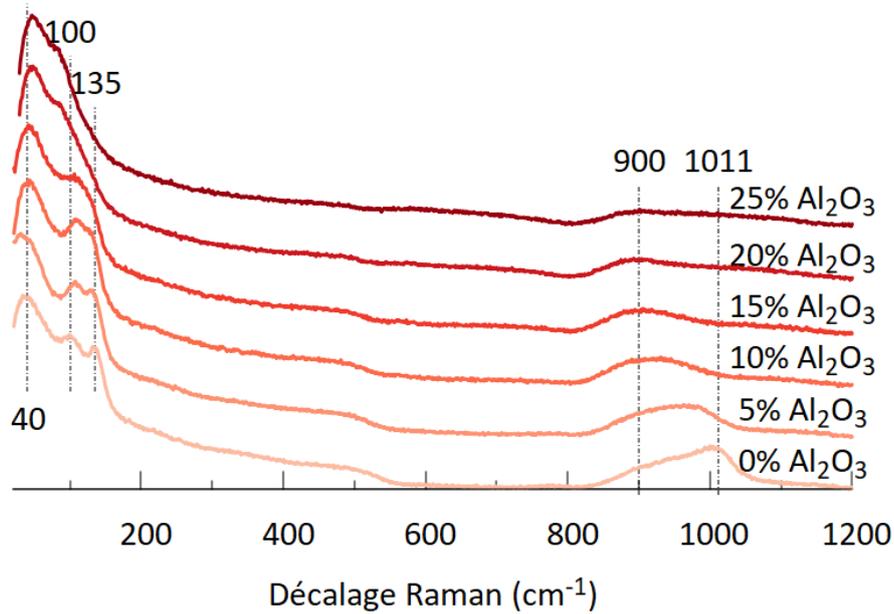

**Figure 3.23.** *Evolution des spectres Raman de verres ternaires SiO$_2$-Al$_2$O$_3$-PbO obtenus en ajoutant des teneurs croissantes en Al2O3 (0 - 25 mol%) au verre binaire 60 SiO$_2$ - 40 PbO (mol%) (Roisine 2018).*

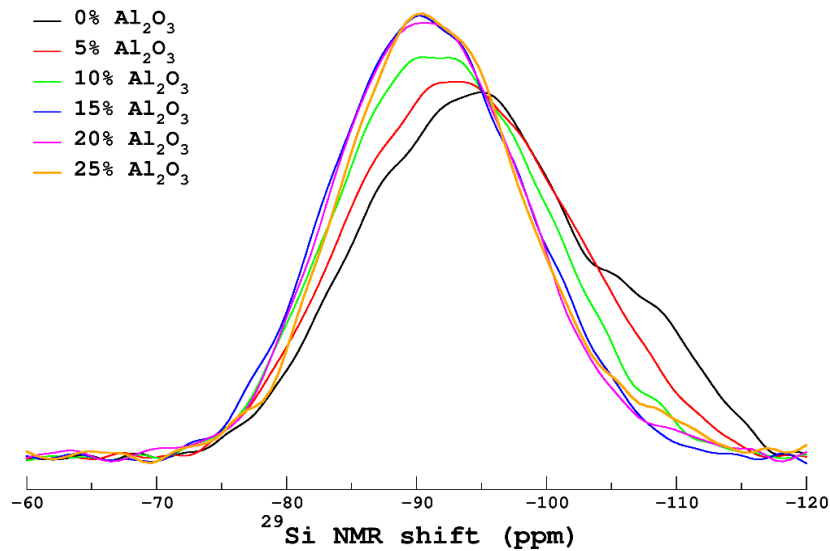

**Figure 3.24.** *Evolution des spectres RMN MAS $^{29}$Si de verres ternaires SiO$_2$-Al$_2$O$_3$-PbO obtenus en ajoutant des teneurs croissantes en Al$_2$O$_3$ (0 - 25 mol%) au verre binaire 60 SiO$_2$ - 40 PbO (mol%). Pour le verre sans Al$_2$O$_3$, les contributions vers -110 et -95 ppm correspondent respectivement aux entités Q$^4$ et Q$^3$ (Roisine 2018).*

En effet, il est connu pour d'autres verres silicatés sans plomb que la présence l'aluminium en coordinence 4 dans l'environnement du silicium (en position de second voisin) conduit à un déplacement des bandes de vibration d'élongation des unités $Q^n$ vers les basses énergies (Rossano et Mysen 2012). L'évolution du signal RMN $^{29}$Si vers les valeurs plus élevées de déplacement chimique avec la teneur en $Al_2O_3$ (figure 3.24) peut s'expliquer de la même façon, le déplacement chimique des unités $Q^n$ augmentant lorsqu'un nombre croissant de tétraèdres $AlO_4^-$ sont présents en seconds voisins des unités $SiO_4$ (Kirkpatrick *et al.* 1985). La région à basse énergie (100 - 200 cm$^{-1}$) des spectres Raman - correspondant aux vibrations Pb-O (Ben Kacem 2017) - évolue également fortement avec la teneur en alumine ajoutée dans les verres binaires $SiO_2$-PbO (figure 3.23). Une forte évolution dans cette région du spectre a également été observée si l'on substitue progressivement PbO par $Al_2O_3$ à teneur constante en $SiO_2$ (Ben Kacem 2017). Cela montre un changement important de l'environnement local des ions $Pb^{2+}$ qui pourrait s'expliquer par la disparition progressive de liaisons assez fortes Pb-O-(Pb,Si) au profit de liaisons Pb-O-Al plus faibles associées au rôle de compensateur de charges des unités $AlO_4^-$ joué par un nombre croissant d'ions $Pb^{2+}$. En effet, l'étude directe de l'environnement de l'aluminium par RMN $^{27}$Al dans ces verres (figure 3.27) montre que celui-ci est toujours très majoritairement présent sous forme $AlO_4^-$ (et donc obligatoirement compensé localement par les ions $Pb^{2+}$ qui sont les seuls cations susceptibles de jouer ce rôle dans les verres ternaires), avec cependant la contribution croissante mais toujours minoritaire d'unités $AlO_5$ mises en évidence par RMN $^{27}$Al MQMAS pour les teneurs en alumine les plus élevées (> 15 mol% $Al_2O_3$) (Roisine 2018).

En ce qui concerne les atomes d'oxygène de la structure, l'étude par RMN $^{17}$O de verres ternaires enrichis en $^{17}$O confirme également une forte évolution de l'environnement du plomb suite à l'ajout d'alumine dans les verres binaires $SiO_2$-PbO (figures 3.25 et 3.26). En effet, on observe une forte diminution de la contribution des liaisons Si-O-Pb au profit de liaisons Si-O-Al montrant ainsi une forte repolymérisation du réseau vitreux (forte diminution de la proportion d'atomes d'oxygène non-pontants). De plus, la faible quantité de liaisons Pb-O-Pb présentes dans le verre binaire semble disparaître dans le verre ternaire (figure 3.25), ce qui tendrait à indiquer une meilleure dispersion du plomb dans le réseau aluminosilicaté en accord avec les résultats obtenus par spectroscopie Raman (région des spectres à basse fréquence).

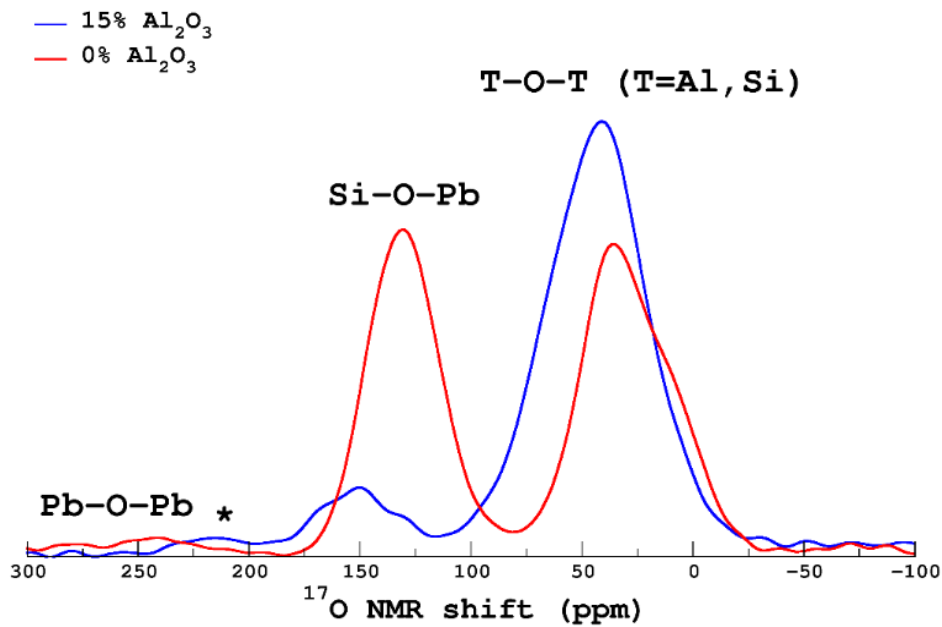

**Figure 3.25.** *Comparaison des spectres RMN MAS $^{17}O$ des verres 60 $SiO_2$ - 40 PbO et 51 $SiO_2$ - 34 PbO - 15 $Al_2O_3$ (mol%) enrichis en $^{17}O$ (*: bande de rotation) (Roisine 2018).*

Cependant, des études complémentaires réalisées sur ces mêmes verres par diffusion des rayons X n'ont pas mis en évidence d'évolution significative de la distance moyenne Pb-Pb lors de l'ajout d'alumine à des verres binaires $SiO_2$-PbO (Roisine 2018). D'après les résultats structuraux concernant la stéréochimie des ions $Pb^{2+}$ dans les phases cristallines riches en $Al_2O_3$ telle que $PbAl_2Si_2O_8$ (section 3.5.2.1) pour laquelle les paires libres $6s^2$ sont stéréochimiquement inactives, on peut penser qu'il en est de même dans les verres ternaires $SiO_2$-$Al_2O_3$-PbO au moins pour les ions $Pb^{2+}$ jouant le rôle de compensateur de charges des unités $AlO_4^-$, les ions $Pb^{2+}$ excédentaires conservent vraisemblablement leur paire $6s^2$ active comme dans les verres binaires $SiO_2$-PbO.

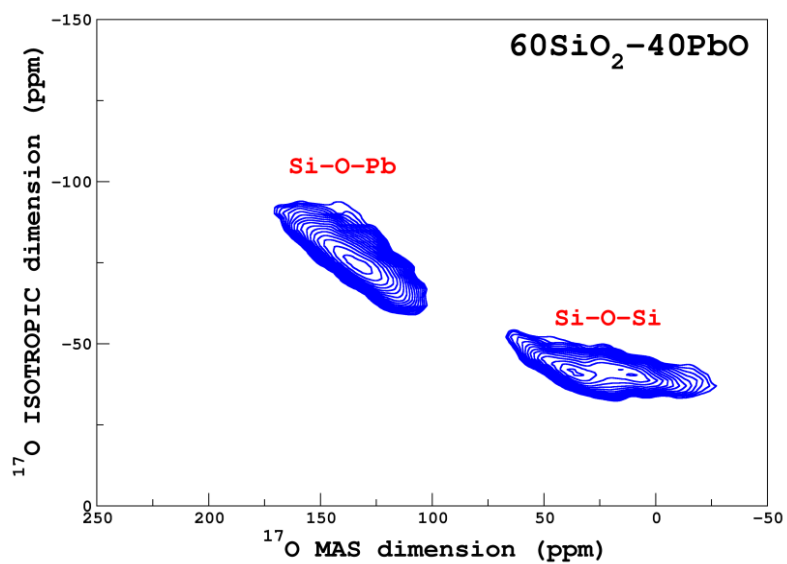

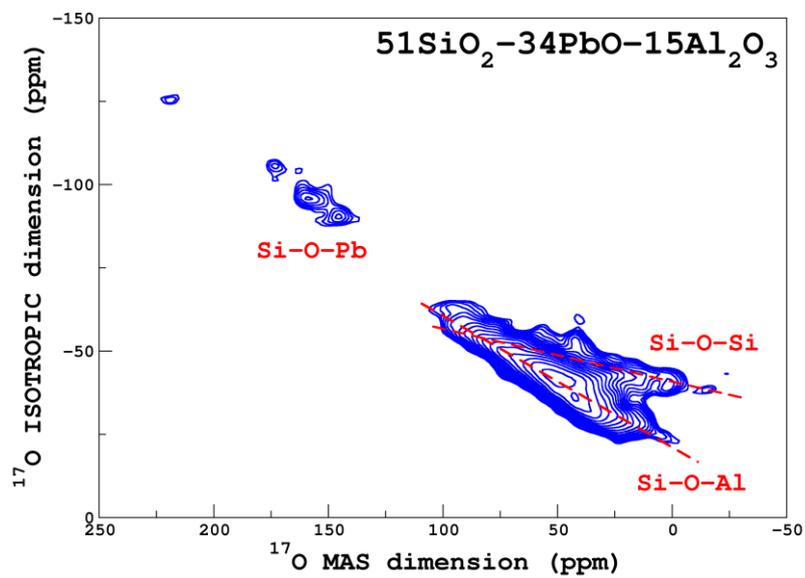

**Figure 3.26.** *Spectres RMN MQ-MAS $^{17}$O des verres 60 SiO$_2$ - 40 PbO et 51 SiO$_2$ - 34 PbO - 15 Al$_2$O$_3$ en mol% enrichis en $^{17}$O (Roisine 2018).*

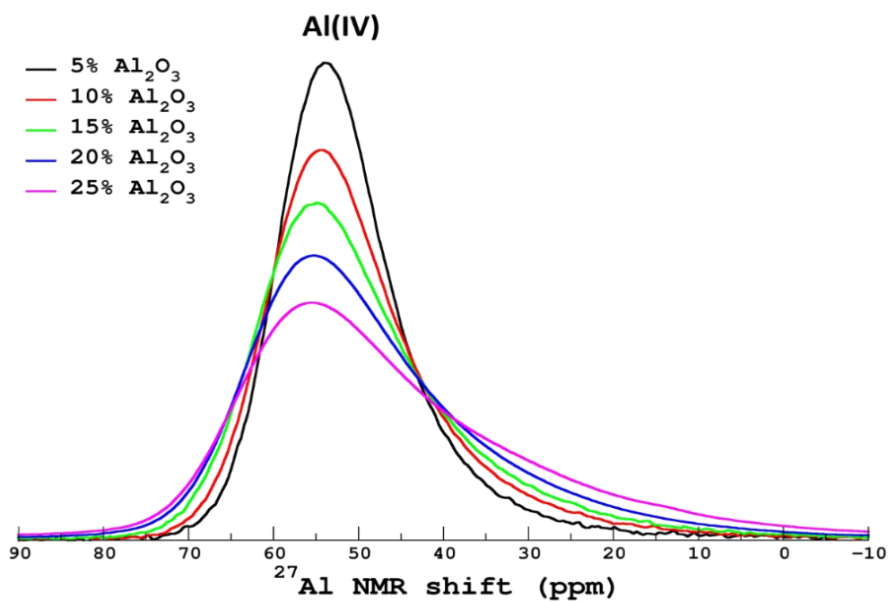

**Figure 3.27.** *Evolution des spectres RMN MAS $^{27}$Al des verres ternaires $SiO_2$-$Al_2O_3$-PbO obtenus en ajoutant des teneurs croissantes en $Al_2O_3$ (0 - 25 mol%) au verre binaire 60 $SiO_2$ - 40 PbO (mol%). L'élargissement et le déplacement du signal vers les valeurs plus élevées de déplacement chimique sont dus respectivement à une augmentation de la constante de couplage quadripolaire $C_Q$ et à une augmentation du déplacement chimique isotrope $\delta_{iso}$ de $^{27}$Al, traduisant une forte évolution de la distorsion des sites $AlO_4^-$ lorsque la teneur en $Al_2O_3$ augmente (Roisine 2018).*

## 3.6. Conclusion

En raison de sa forte polarisabilité et de sa configuration électronique particulière (paire libre $6s^2$) l'ion $Pb^{2+}$ présente une stéréochimie singulière qui conduit à l'obtention de verres binaires silicatés $SiO_2$-PbO pouvant renfermer des teneurs très élevées en PbO, bien plus élevées en particulier que la quantité d'oxydes alcalins $R_2O$ qu'il est possible d'introduire dans les verres binaires de silicates alcalins ($SiO_2$-$R_2O$). Bien que les premières études structurales menées sur les verres binaires de silicate de plomb remontent à plus de 80 ans, des interrogations demeurent toujours sur le rôle structural de PbO dans les verres silicatés et sur l'évolution de la répartition des ions $Pb^{2+}$ en fonction de la concentration en plomb. Cependant, la mise en œuvre de nombreuses méthodes d'investigation couplées à des méthodes de simulation numérique ont conduit à de nombreuses publications et à une connaissance de plus en plus précise de la structure de ces verres. Ainsi, pour les verres binaires du système $SiO_2$-PbO, il apparaît que - comme pour les phases cristallines PbO et les silicates de plomb - les ions $Pb^{2+}$ sont localisés dans des sites très asymétriques de faible coordinence (3,4) avec leur paire libre $6s^2$ stéréochimiquement active. Bien que l'ajout de PbO conduise à une dépolymérisation du réseau silicaté, le rôle structural des ions $Pb^{2+}$ est très différent de celui des ions bivalents comme les ions alcalino-terreux. En effet, d'une part des liaisons bien plus fortes s'établissent dès les faibles teneurs en PbO entre les ions $Pb^{2+}$ et les atomes d'oxygène non-pontants du réseau silicaté, et d'autre part pour les teneurs élevées en PbO une proportion croissante d'ions $O^{2-}$ se trouvent connectés uniquement aux ions $Pb^{2+}$ formant progressivement un réseau plombifère de plus en plus étendu. Mais malgré tout, même pour les teneurs en PbO les plus fortes, une proportion significative d'unités $SiO_4$ restent connectées entre elles pouvant ainsi contribuer à stabiliser le réseau plombifère vis-à-vis de la cristallisation. En revanche, les études structurales réalisées sur d'autres systèmes riches en plomb sont bien moins nombreuses mais sont cependant riches d'enseignements. Ainsi, dans les verres de silicates alcalins, pour les faibles teneurs en plomb les ions $Pb^{2+}$ tendent à s'incorporer dans des zones renfermant des ions alcalins où ils s'associent aux atomes d'oxygène non-pontants, alors que pour les teneurs élevées en PbO apparaissent des connexions Pb-O-Pb, signature de la formation d'un réseau plombifère d'extension croissante. Au sein de ces verres, les ions $Pb^{2+}$ conserveraient leur paire libres $6s^2$ stéréochimiquement active et seraient localisés dans des sites très asymétriques par analogie aux phases cristallines. En revanche, dans le cas des verres d'aluminosilicate de plomb, les ions $Pb^{2+}$ tendent à jouer le rôle de compensateur de charge des unités $AlO_4^-$ et perdent vraisemblablement l'activité stéréochimique de leur paire libre $6s^2$. Toutefois, des interrogations demeurent sur l'évolution du réseau plombifère en fonction de la teneur en $Al_2O_3$ dans ces verres.

## 3.7. Bibliographie

Alderman, O.L.G., Hannon, A.C., Holland, D., Feller, S., Lehr, G., Vitale, A.J., Hoppe, U., Zimmerman, M.V., Watenphule, A.(2013). Lone-pair distribution and plumbite network formation in high lead silicate glass, $80PbO.20SiO_2$. *Phys. Chem. Chem. Phys.*, 15, 8506-8519.

Alderman, O.L.G. (2013). The structure of vitreous binary oxides: silicate, germanate and plumbite networks. Thèse de doctorat de l'Université de Warwick.

Andersen, O. (1919). The volatilization of lead oxide from lead silicate melts. *J. Am. Ceram. Soc.*, 2, 784-789.

Angeli, F., Jollivet, P., Charpentier, T., Fournier, M., Gin, S. (2016). Structure and chemical durability of lead crystal glass. *Environ. Sci. Technol.*, 50, 11549-11558.

Ashour, G.M., Shehata, A.M. (1972). Study of crystallization of chromium oxide in glass (chrome aventurine). *Trans. Indian Ceram. Soc.*, 31, 136-143.

Bair, C.J.(1936). The constitution of lead oxide-silica glasses: I, atomic arrangement. *J. Am. Ceram. Soc.*, 19, 339-347.

Bassi, G., Lajzerowicz-Bonneteau, J. (1965). Structure du pyrogermanate de plomb potassium $K_2Pb_2Ge_2O_7$. *Bull. Soc. Fr. Minéral. Crist.*, 88, 342-344.

Ben Kacem, I. (2017). Du cristal au plomb jusqu'aux déchets domestiques : rôle du plomb dans les verres et les vitrocéramiques : étude des systèmes $PbO-SiO_2$, $PbO-CaO-SiO_2$ et $PbO-Al_2O_3-SiO_2$. Thèse de doctorat de l'Université Paris-Est.

Ben Kacem, I., Gautron, L., Coillot, D., Neuville, D.R.(2017). Structure and properties of lead silicate glasses and melts. *Chem. Geol.*, 461, 104-114.

Benna, P., Tribaudino, M., Bruno, E. (1996). The structure of ordered and disordered lead feldspar ($PbAl_2Si_2O_8$). *Am. Mineral.*, 81, 1337-1343.

Bessada, C., Massiot, D., Coutures, J., Douy, A., Coutures, J.-P., Taulelle, F. (1994). $^{29}$Si MAS-NMR in lead silicates. *J. Non-Cryst. Solids*, 168, 76-85.

Bhargavi, K., Reddy, M.S., Rao, P.R., Rao, N.N., Rao, M.S., Kumar, V.R., Veeraiah, N.(2012). The structural influence of aluminium ions on emission characteristics of $Sm^{3+}$ ions in lead aluminium silicate glass system., *Mater. Res. Bull.*, 47, 267-273.

Bhargavi, K., Rao, M.S., Veeraiah, N., Sanyal, B., Gandhi, Y., Baskaran, G.S.(2014). Thermal Luminescence of $Ho_2O_3-PbO-Al_2O_3-SiO_2$ Glasses Exposed to Gamma Radiation. *Int. J. Appl. Glass Sci.*, 6, 128-136.

Bhargavi, K., Sanyal, B., Rao, M.S., Kumar, V., Gandhi, Y., Baskaran, G.S., Veeraiah, N.(2015). γ-Ray induced thermoluminescence characteristics of the $PbO-Al_2O_3-SiO_2:Pr^{3+}$ glass system. *J. Lumin.*, 161, 417-421.

Boucher, M.L. Peacor, D.R.(1968). The crystal structure of alamosite, $PbSiO_3$. *Z. Kristallogr.*, 126, 98-111.

Brandes, R., Hoppe, R. (1994). $Li_{10}Si_2Pb^{II}O_{10} = Li_{20}[(SiO_4)_4(OPbO_2PbO)]$ –Daserste „gemischte" Silicat-Plumbat(II), Z. Anorg. Allg. Chem., 620, 2026-2032.

Brown, I.D., Altermatt, D. (1985). Bond-valence parameters obtained from a systematic analysis of the inorganic crystal structure database. Acta Cryst., B41, 244-247.

Calvert, P.D., Shaw, R.R.(1970). Liquidus behavior in the silica-rich region of the system $PbO-SiO_2$. *J. Am. Ceram. Soc.*, 53, 350-352.

Carr, S.M., Subramanian, K.N.(1982a). Spherulitic crystal growth in $P_2O_5$-nucleated lead silicate glasses. *J. Non-Cryst. Solids*, 60, 307-312.

Carr, S.M., Subramanian, K.N. (1982b). Effect of $P_2O_5$ on the crystallization of lead silicate glasses. *J. Am. Ceram. Soc.*, 65, 346-349.

Charles, R.J.(1966). Metastable Liquid Immiscibility in Alkali Metal Oxide-Silica Systems. *J. Am. Ceram. Soc.*, 49, 55-62.

Charpentier, T., Menziani, M.C., Pedone, A.(2013). Computational simulations of solid state NMR spectra: a new era in structure determination of oxide glasses. *RCS Adv.*, 3, 10550-10578.

Chen, S., Zhao, B., Hayes, P. C., Jak, E.(2001). Experimental study of phase equilibria in the $PbO-Al_2O_3-SiO_2$ system. *Metal. Mater. Trans. B*, 32B, 997-1005.

Chevrier, G., Giester, G., Hege,r G., Jarosch, D., Wildner, M., Zemann, J. (1992). Neutron single-crystal refinement of cerussite, $PbCO_3$, and comparison with other aragonite-type carbonates. *Z. Kristallogr.*, 199, 67-74.

Dalby, K.N., Wayne Nesbitt, H., Zakaznova-Herzog, V.P., King, P.L.(2007). Resolution of bridging signals from O 1s spectra of silicate glasses using XPS: Implications for O and Si speciation. *Geochem. Cosmochim. Acta*, 71, 4297-4313.

Dayanand, C., Bhikshamaiah, G., Jaya Tyagaraju, V., Salagram, M., Krishna Murthy, A.S. R. (1996). Structural investigations of phosphate glasses: a detailed infrared study of the $x(PbO)-(1-x)P_2O_5$ vitreous system. *J. Mater. Sci.*, 31, 1945-1967.

Dent Glasser, L.S., Howie, R.A., Smart, R.M.(1981). The structure of lead orthosilicate, $2PbO.SiO_2$. *Acta Cryst.*, *B* 37, 303-306.

Dietzel, A.(1948). Structure and properties of glass. *Glastechn. Ber.*, 22, 41-50.

Dimitrov, V., Komatsu, T.(2010). An interpretation of optical properties of oxides and oxide glasses in terms of the electronic polarizability and average single bond strength (review). *J. Univ. Chem. Technol. Metall.*, 45, 219-250.

El-Egili, K., Doweidar, H., Moustafa, Y.M., Abbas, I.(2003). Structure and some physical properties of $PbO-P_2O_5$ glasses. *Physica B*, 339, 237-245.

Ellis, E., Johnson, D.W., Breeze, A., Magee, P.M., Perkins, P.G.(1979). The electronic band structure and optical properties of oxide glasses. *Philos. Mag. B*, 40, 125-137.

El Shamy, T.M., Taki-Eldin, H.D.(1974). The chemical durability of $PbO-SiO_2$ glasses. *Glass Technol.*, 15, 48-52.

Faivre, R., Weiss, R. (1963). Le plomb. Dans *Nouveau traité de chimie minérale, Tome 8*, Pascal, P. (dir). Masson, Paris.

Fajans, K., Kreidl, N.J. (1948). Stability of lead glasses and polarization of ions. *J. Am. Ceram. Soc.*, 31, 105-114.

Fayon, F., Farnan, I., Bessada, C., Coutures, J., Massiot, D., Coutures, J.-P. (1997). Empirical correlations between $^{207}$Pb NMR chemical shifts and structure in solids. *J. Am. Chem. Soc.*, 119, 6837-6843.

Fayon, F. (1998). Caractérisation du rôle structural du plomb dans des matrices vitreuses : approche par RMN haute résolution solide mono et bidimensionnelle. Thèse de Doctorat de l'Université d'Orléans.

Fayon, F., Bessada, C., Massiot, D., Farnan, I., Coutures, J.-P. (1998). $^{29}$Si and $^{207}$Pb NMR study of local order in lead silicate glasses. *J. Non-Cryst. Solids*, 232-234, 403-408.

Fayon, F., Landron, C., Sakurai, K., Bessada, C., Massiot, D. (1999). $Pb^{2+}$ environment in lead silicate glasses probed by Pb-L$_{III}$ edge XAFS and $^{207}$Pb NMR. *J. Non-Cryst. Solids*, 243, 39-44.

Feller, S., Riley, A., Edwards, T., Croskrey, J., Schue, A., Liss, D., Stentz, D., Blair, S., Kelley, M., Smith, G., Singleton, S., Affatigato, M., Holland, D., Smith, M.E., Kamitsos, E.I., Varsamis, C.P.E., Ioannou, E. (2010). A multispectroscopic structural study of lead silicate glasses over an extended range of compositions. *J. Non-Cryst. Solids*, 356, 304-313.

Flood, H., Förland, T. (1947). The acidic and basic properties of oxides. *Acta Chem. Scand.*, 1, 592-604.

Freed, R.L., Peacor, D.R. (1969). Determination and refinement of the crystal structure of margarosanite $PbCa_2Si_3O_9$. *Z. Kristallogr.*, 128, 213-228.

Frieser, R.J. (1975). A review of solder glasses. *Electro. Comp. Sci. Tech.*, 2, 163-199.

Furukawa, T., Brawer, S.A., White, W.B. (1978). The structure of lead silicate glasses determined by vibrational spectroscopy. *J. Mater. Sci.*, 14, 268-282.

Gee, I.A., Holland, D., McConville, C.F. (2001). Atomic environments in binary lead silicate and ternary alkali lead silicate glasses. *Phys. Chem. Glasses*, 42, 339-348.

Geller, R.F., Creamer, A.S., Bunting, E.N. (1934). The system $PbO-SiO_2$. *J. Res. Natl. Bur. Stand.*, 13, 237-244.

Greenough, D.R., Dentschuk, P., Palmer, S.B. (1981). Thermal expansion of lead silicate glasses. *J. Mater. Sci.*, 19, 599-603.

Gillespie, R.J., Nyholm, R.S.Q. (1957). Inorganic stereochemistry. *Rev. Chem. Soc.*, 11, 339-380.

Gourlaouen, C., Parisel, O. (2007). Is an electronic shield at the molecular origin of lead poisoning? A computational modeling experiment. *Angew. Chem.*, 119, 559-562.

Hirota, K., Hasegawa, Y.T.(1981). Phase Relations in the System PbO-PbSiO$_3$.*Bull. Chem. Soc. Jpn.*,54, 754-756.

Hoppe, U., Kranold, R., Ghosh, A., Landron, C., Neuefeind, J., Jovari, P.(2003). Environments of lead cations in oxide glasses probed by X-ray diffraction. *J. Non-Cryst. Solids*, 328, 146-156.

Imaoka, M., Yamazaki, T.(1963). Studies of the glass-formation range silicate systems. Investigations on the glass-formation range 2. *J. Ceram. Assoc. Jpn.*, 71, 215-223.

Jak,E., Hayes P.C., DegterovS., Pelton A.D., Wu P. (1997). Thermodynamic optimization of the systems PbO- SiO$_2$, PbO-ZnO, ZnO-SiO$_2$ and PbO-ZnO-SiO$_2$. *Metall. Mater. Trans. B.*, 28, 1011-1018.

Kantor, P., Revcolevschi, A., Collongues, R. (1973). Preparation of iron sesquioxide glasses by ultra-fast quenching from the melt ("splat-cooling"). *J. Mater. Sci.*, 8, 1359-1361.

Kato, K. (1982). Die Kristallstruktur des Bleisilicats Pb$_{11}$Si$_3$O$_{17}$. *Acta Cryst. B*, 38, 57-62.

Kaur, A., Khanna, A., Singla, S., Dixi, A., Kothiyal, G.P., Krishnan, K., Aggarwal, S.K., Sath, V., González, F., González-Barriuso, M. (2013). Structure–property correlations in lead silicate glasses and crystalline phases. *Phase Transit.*, 86, 759-777.

Kirkpatrick, R.J., Smith, K.A., Schramm, S., Turner, G., Yang, W.-H.(1985). Solid-state nuclear magnetic resonance spectroscopy of minerals.*Ann. Rev. Earth Planet. Sci.*, 13, 29-47.

Kohara, S., Ohno, H., Takata, M., Usuki, T., Morita, H., Suzuya, K., Akola, J., Pusztai, L.(2010). Lead silicate glasses: Binary network-former glasses with large amounts of free volume. *Phys. Rev. B*, 82, 134209.

Korson, L., Drost-Hansen,W., Millero, F. J.(1969). Viscosity of water at various temperatures. *J. Phys. Chem.*, 73, 34-39.

Lee, S.K., Kim, E.J.(2015). Probing metal-bridging oxygen and configurational disorder in amorphous lead silicates: insights from $^{17}$O solid-state nuclear magnetic resonance. *J. Phys. Chem. C*, 119, 748-756.

Lippmaa, E., Samoson, A., Mägi, M., Teeäär, R., Schraml, J., Götz, J. (1982). High resolution $^{29}$Si NMR study of the structure and devitrification of lead-silicate glasses. *J. Non-Cryst. Solids*, 50, 215-218.

Maekawa, H., Maekawa, T., Kawamura, K., Yokokawa, T.(1991). The structural groups of alkali silicate glasses determined from $^{29}$Si MAS-NMR. *J. Non-Cryst. Solids*, 127, 53-64.

Martlew, D. (2005). Viscosity of molten glasses. Dans *Properties of glass-forming melts*. Pye, L.D., Montenero, A., Joseph, I. (dir). CRC Press, Boca Raton, 119-120.

McMurdie, H.F. (1941). $K_2Pb_2Si_2O_7$. *J. Res. Natl. Bur. Stand.*, 26, 489.

Mizuno, M., Takahashi, M., Takaishi, T., Yoko, T. (2005). Leaching of lead and connectivity of plumbate networks in lead silicate glasses. *J. Am. Ceram. Soc.*, 88, 2908-2912.

Morikawa, H., Jegoudez, J., Mazières, C., Revcolevschi, A. (1981). Glass forming ability and crystallization in the $PbO-Al_2O_3$ system. *J. Non-Cryst. Solids*, 44, 107-112.

Mydlar, M.F., Kreidl, N.J., Hendren, J.K., Clayton, G.T. (1970). X-ray diffraction study of lead silicate glasses. *Phys. Chem. Glasses*, 11, 196-204.

Mylyanych, L., Sheredko, M.A., Melnyk, S.K. (1999). Study of glass structures and crystalline phases in the $PbO-Al_2O_3-SiO_2$ system. *J. Anal. At. Spectrom.*, 14, 513-521.

Mysen, B., Richet, P. (2005). *Silicate glasses and melts*. Elsevier Sciences, Amsterdam.

Neiman, T.S., Yinnon, H., Uhlmann, D.R. (1982). Crystallization kinetics of lead metasilicate. *J. Non-Cryst. Solids*, 48, 393-403.

O'Shaughnessy, C., Henderson, G.S., Wayne Nesbitt, H., Bancroft, G.M., Neuville, D.R. (2020). The influence of modifier cations on the Raman stretching modes of $Q_n$ species in alkali silicate glasses. *J. Am. Ceram. Soc.*, 103, 3991-4001.

Palomar, T., Mosa, J., Aparicio, M. (2020). Hydrolytic resistance of $K_2O-PbO-SiO_2$ glasses in aqueous and high-humidity environments. *J. Am. Ceram. Soc.*, 103, 5248-5258.

Pannhorst, W., Lohn, J. (1970). Zur Kristallstruktur von Strontianit, $SrCO_3$. *Z. Kristallogr.*, 131, 455-459.

Patrick, L. (2006). Lead toxicity, a review of the literature. Part 1: Exposure, evaluation, and treatment. *Altern. Med. Rev.*, 11, 2-22.

Pena, R.B., Sampaio, D.V., Lancelotti, R.F., Cunha, T.R., Zanotto, E.D., Pizani, P.S. (2020). In-situ Raman spectroscopy unveils metastable crystallization in lead metasilicate glass. *J. Non-Cryst. Solids*, 546, 120254.

Petter, W., Harnik, A.B., Keppler, U. (1971). Die Kristallstruktur von Blei-Barysilit, $Pb_3Si_2O_7$. *Z. Kristallogr.*, 133, 445-458.

Prabakar, S., Rao, K. J. (1991). MAS NMR studies of ternary silicate glasses. *Phil. Mag. B*, 64, 401-411.

Pye, L.D. (1988). Aluminate Glasses - A Review. *Proc. SPIE 0929, Infrared Optical Materials IV*, 149-156.

Rabinovich, E. M.(1976). Lead in glasses. *J. Mater. Sci.*, 11, 925-948.

Rao, M.S., Sudarsan, V., Brik, M.G., Bhargavi, K., Rao, Ch.S., Gandhi, Y., Veeraiah, N.(2013). The de-clustering influence of aluminum ions on the emission features of $Nd^{3+}$ ions in $PbO$–$SiO_2$ glasses. *Opt. Commun.*, 298-299, 135-140.

Rao, M.S., Sanyal, B., Bhargavi, K., Vijay, R., Kityk, I.V., Veeraiah, N.(2014). Influence of induced structural changes on thermoluminescence characteristics of γ-ray irradiated $PbO$–$Al_2O_3$–$SiO_2$: $Dy^{3+}$ glasses. *J. Mol. Struct.*, 1073, 174-180.

Roisine, G. (2018). Céramiques glaçurées de Bernard Palissy : A la recherche des secrets d'un maître de la Renaissance. Thèse de Doctorat de l'Université Paris Sciences et Lettres (PSL).

Ronay, C., Seff, K. (1985). Crystal structures of $Pb_6$-A and $Pb_9(OH)_8(H_2O)_3$-A. Zeolite A ion exchanged with $Pb^{2+}$ at pH 4.3 and 6.0 and evacuated. *J. Phys. Chem.*, 89, 1965-1970.

Rossano, S., Mysen, B.(2012). Raman spectroscopy applied to earth sciences and cultural heritage. Dans *EMU Notes in Mineralogy, Vol. 12*, Dubessy, J., Caumon, M.-C., Rull, F. (dir). Eotvos University Press, Budapest, 319-364.

Roy, B.N. (1990). Infrared spectroscopy of lead and alkaline-earth aluminosilicate glasses. *J. Am. Ceram. Soc.*, 73, 846-855.

Rybicki, J., Rybicka, A., Witkowska, A., Bergmanski, G., Di Cicco, A., Minicucci, M., Mancini G.(2001). The structure of lead-silicate glasses: molecular dynamics and EXAFS studies. *J. Phys. Condens. Matter*, 13, 9781-9797.

Saini, A., Khanna, A., Michaelis, V.K., Kroeker, S., González, F., Hernández, D. (2009). Structure–property correlations in lead borate and borosilicate glasses doped with aluminum oxide. *J. Non-Cryst. Solids*, 355, 2323-2332.

Sawvel, A.M., Chinn, S.C., Bourcier, W.L., Maxwell, R.S.(2005). Local structure of amorphous $(PbO)_x[(B_2O_3)_{1-z}(Al_2O_3)_z]_y(SiO_2)_y$ dielectric materials by multinuclear solid state NMR. *Chem. Mater.*, 17, 1493-1500.

Scholze, H. (1980). *Le verre. Nature, structure et propriétés*. Institut du verre, Paris.

Seward, T.P., Uhlmann, D.R., Turnbull, D. (1968). Phase separation in the system $BaO$-$SiO_2$. *J. Am. Ceram. Soc.*, 51, 278-285.

Shannon, R.D. (1976). Revised effective ionic radii and systematic studies of interatomic distances in halides and chacogenides. *Acta Crysta*, A32, 751-767.

Shannon, R.D., Fischer, R.X. (2016). Empirical electronic polarizabilities of ions for the prediction and interpretation of refractive indices: Oxides and oxysalts. *Am. Mineral.*, 101, 2288-2016.

Shelby, J.E.(1983). Property/structure relationships in lead silicate glasses. *Glastech. Ber.*, 56, 1057-1062.

Shevchenko, M., Jak, E. (2018). Experimental phase equilibria studies of the PbO–SiO$_2$ system. *J. Am. Ceram. Soc.*, 101, 458-471.

Shrikhande, V.K., Sudarsan, V., Kothiyal, G.P., Kulshreshtha, S.K.(2001). $^{29}$SiMAS NMR and microhardness studies of some lead silicate glasses with and without modifiers. *J. Non-Cryst. Solids*, 283, 18-26.

Shrikhande, V.K., Sudarsan, V., Kothiyal, G.P., Kulshreshtha, S.K.(2007). Photoluminescence and structural studies on Na$_2$O-PbO-SiO$_2$ glasses. *J. Non-Cryst. Solids*, 353, 1341-1345.

Siidra, O.I., Zenko, D.S., Krivovichev, S.V.(2014). Structural complexity of lead silicates: Crystal structure of Pb$_{21}$[Si$_7$O$_{22}$]$_2$[Si$_4$O$_{13}$] and its comparison to hyttsjöite. *Am. Mineral.*, 99, 817-823.

Singh, K.J., Kaur, S., Kaundal, R.S.(2014).Comparative study of gamma ray shielding and some properties of PbO-SiO$_2$-Al$_2$O$_3$ and Bi$_2$O$_3$-SiO$_2$-Al$_2$O$_3$ glass systems. *Radiat. Phys. Chem.*, 96, 153-157.

Sinn, E.(1979). Measurement of the shear viscosity in garnet high temperature solutions by a cylindrical rotation viscometer. *Krist. Tech.* 14, 117-127.

Smart, R.M., Glasser, F.P.(1974). Compound formation and phase equilibria in the system PbO-SiO$_2$. *J. Am. Ceram. Soc.*, 57, 378-382.

Smets, B.M.J., Lommen, T.P.A.(1982). The structure of glasses and crystalline compounds in the system PbO-SiO$_2$, studied by X-ray photoelectron spectroscopy. *J. Non-Cryst. Solids*, 48, 423-430.

Sorokina, M.F., Kanunnikova, O.M., Gil'mutdinov, F.Z., Kozhevnikov, V.I. (1996). X-ray electron spectroscopy investigation of the structure of double lead silicate glasses. *Glass Ceram.*, 53, 11-13.

Stols-Witlox, M. (2011).*The heaviest and the whitest: lead white quality in north western European documentary sources, 1400-1900*. Archetype Editions, Londres.

Suzuya, K., Kohara, S., Ohno, H.(1999). A reverse Monte Carlo study of lead metasilicate glass. *Jpn. J. Appl. Phys.*, 38 (Suppl. 38-1), 144-147.

Takaishi, T., Takahashi, M., Jin, J., Uchino, T., Yoko, T. (2005). Structural study on PbO-SiO$_2$ glasses by X-ray and neutron diffraction and $^{29}$Si MAS NMR measurements. *J. Am. Ceram. Soc.*, 88, 1591-1596.

Takamori, T. (1979). Solder glasses. Dans *Treatise on Materials Science and Technology vol. 17 Glass II*, Tomozawa, M. Doremus, R.H. (dir.), Academic Press, New York, 173-255.

Tite, M.S., Freestone, I.C., Mason, R., Molera, J., Vendrell-Saz, M., Wood, N. (1998). Review article. Lead glazes in antiquity - methods of production and reasons for use. Archaeometry, 40, 241-260.

Vogel, W. (1994). *Glass chemistry*. Springer-Verlag (2$^{nd}$ Edition), Berlin, Heidelberg.

Volf, M.O. (1984). *Chemical approach to glass*. Glass Science and Technology 7, Elsevier, Amsterdam, New York, 443-464.

Waghmare, U.V., Spaldin, N.A., Kandpal, H.C., Ram Seshadri, (2003). First-principle indicators of metallicity and cation off-centricity in the IV-VI rocksalt chalcogenides of divalent Ge, Sn and Pb. *Phys. Rev. B*, 67, 125111.

Walsh, A., Watson, G.W. (2005). The origin of the stereochemically active Pb(II) lone pair: DFT calculations on PbO and PbS. *J. Solid State Chem.*, 178, 1422-1428.

Wang, P.W., Zhang, L. (1996). Structural role of lead in lead silicate glasses derived from XPS spectra. *J. Non-Cryst. Solids*, 194, 129-134.

Warren, B.E., Loring, A.D. (1935). X-ray diffraction study of the structure of soda-silica glass. *J. Am. Ceram. Soc.*, 18, 269-276.

Watson, G.W., Parker, S.G., Kresse, G. (1999). Ab initio calculation of the origin of the distortion of alpha-PbO. *Phys. Rev. B*, 59, 8481-8486.

Welcomme, E., Walter, P., Bleuet, P., Hodeau, J.-L., Dooryhee, E., Martinetto, P., Menu, M. (2007). Classification of lead white pigments using synchrotron radiation micro X-ray diffraction. *Appl. Phys. A*, 89, 825-832.

Yeom, Y.H., Kim, Y., Seff, K. (1997). Crystal structure of zeolite X exchanged with Pb(II) at pH 6.0 and dehydrated: $(Pb^{4+})_{14}(Pb^{2+})_{18}(Pb_4O_4)_8Si_{100}Al_{92}O_{384}$, *J. Phys. Chem. B*, 101, 5314-5318.

Yoshihoto, M., Soga, N. (1987). The effect of composition on crack propagation in lead silicate glasses. *J. Non-Cryst.* Solids, 95-96, 1039-1046.

Zachariasen, W.H. (1932). The atomic arrangement in glass. *J. Am. Chem. Soc.*, 54, 3841-3851.